\documentclass[aps,pra,twocolumn,amsmath,amssymb,superscriptaddress,showpacs]{revtex4-1}

\usepackage{amsfonts}
\usepackage{xcolor}
\usepackage{amsmath}
\usepackage{amssymb}
\usepackage{amsthm}
\usepackage{fontenc}
\usepackage{graphicx}
\usepackage{xcolor}
\usepackage{textcomp}
\usepackage{epstopdf}
\usepackage{braket}
\usepackage{mathtools}
\usepackage{amsmath}
\usepackage{dcolumn}
\usepackage{multirow}
\usepackage{units}

\begin{document}
\newcommand{\abs}[1]{\lvert#1\rvert}

\title{Electronic fractal patterns  in building Sierpinski-triangles molecular systems}

\author{L.L.Lage}
\affiliation{Instituto de F\' isica, Universidade Federal Fluminense, Niter\' oi, Av. Litor\^ anea sn 24210-340, RJ-Brazil}
\author{A. Latg\' e}
%\email[A. Latg\'e]{andrea.latge@gmail.com}
\affiliation{Instituto de F\' isica, Universidade Federal Fluminense, Niter\' oi, Av. Litor\^ anea sn 24210-340, RJ-Brazil}

\date{\today}

\begin{abstract}

The Sierpinski Triangle (ST) is a fractal mathematical structure that has been used to explore the emergence of flat bands in lattices of different geometries and dimensions in condensed matter.  Here we look into fractal features in the electronic properties of ST flakes and molecular chains simulating experimental synthesized fractal nanostructures. We use a single-orbital tight binding model to study fractal properties of the electronic states and the Landauer formalism to explore transport responses of the quasi 1D molecular chains. The self-similarity of the energy states are found comparing different ST orders and also amplifying the energy ranges investigated, for both flakes and quasi-1D systems. In particular, the results for the local density of states of the theoretical molecular chains proposed here exhibit quite similar spatial charge distribution of experimental STM reports. The analysis of transport response of such all-carbon fractal molecular chains can be used as a guide to propose a variety of architecture in the synthesis of real new molecular chains.

\end{abstract}

\maketitle

\section{Introduction}
The physical properties of confined electronic structures have been widely investigated, mainly thanks to the possibility of tuning electronic states and its transport responses. Nanostructured systems emerge in different  shapes and dimensions and further than being only dreamed theoretical idealization are being synthesized making use of multiple bottom-up and up-bottom growth processes \cite{Review2020,Kagome2022}. In particular, systems exhibiting dispersionless states in the electronic structure, named as flat bands have been pursued with the help of different strategies, such as controlling the angle of twisted bilayers \cite{Twisted2021,Cao2018}, switching on electric and magnetic fields \cite{LandauNature2020,PhysRevLett.123.096802}, and also engineering the growth of buckled graphene superlattices \cite{Andrei2020}.

Recently, the fractal dimension opened a new door in the study of the electronic properties in confined systems \cite{nature}. The Sierpinski Triangle (ST) is a fractal mathematical structure that was used physically as a quantum corral to confine electrons. 
Transport responses were also explored in ST lattice models indicating a new scenario for correlated topological systems in fractal dimensions \cite{Biplab2020}. Despite the fact ST is represented in a bidimensional form, it has a fractal dimension, called Hausdorff dimension (D) given by %$D= log(3)/log(2) \approx  1.585$
$D\approx1.585$.

\begin{figure}[!h]
    \centering
    \includegraphics[width=9cm]{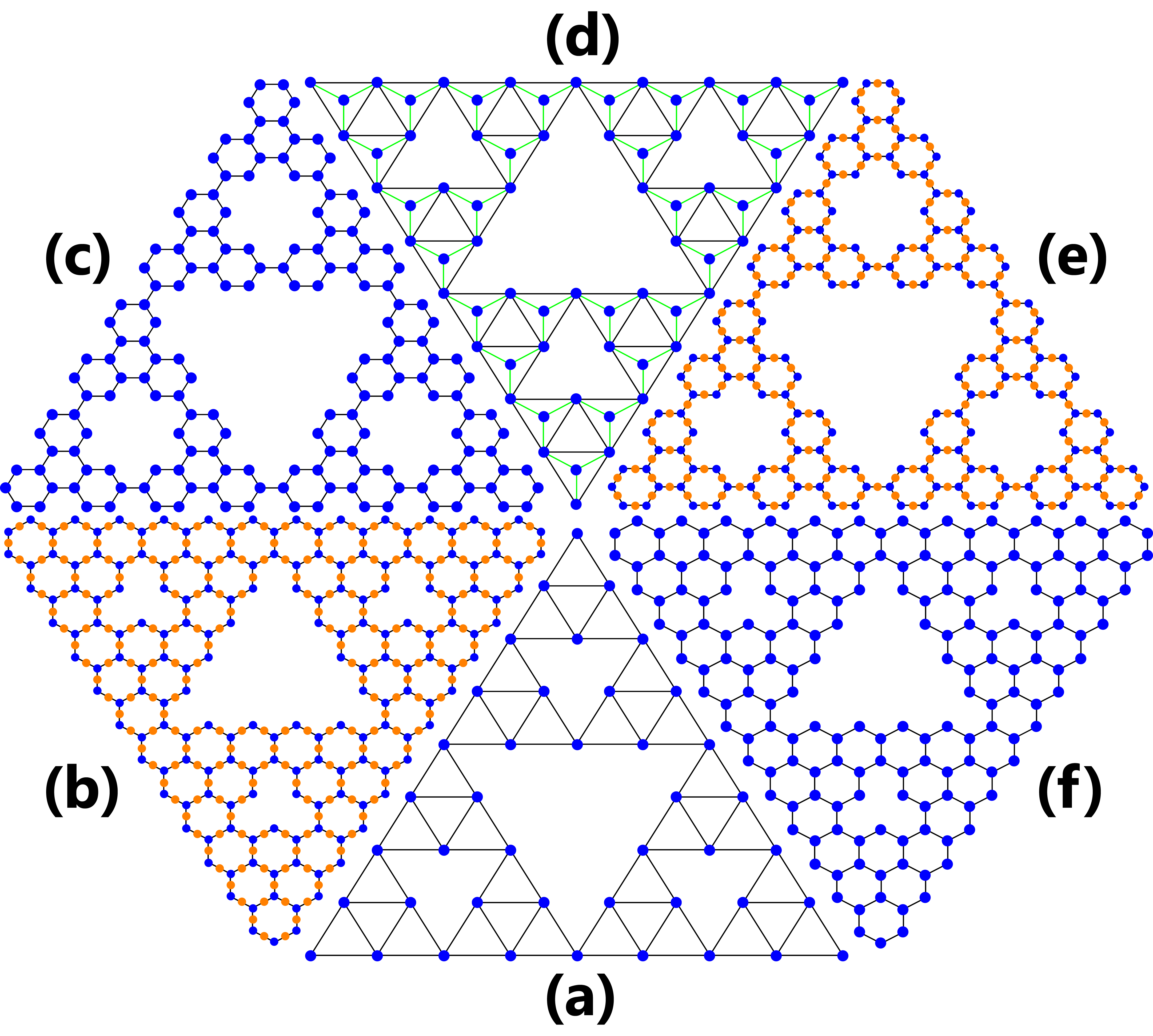}
   \caption{Variations of G(3) Sierpinski Triangle flakes: (a) Conventional ST, (b) Zigzag-Kagomé-Honeycomb ST (Z-KHST), (c) Armchair-Graphene ST (A-GST), (d) Hexagonal ST (HST), (e) Armchair-Kagomé-Honeycomb ST (A-KHST), and (f) Zigzag-Graphene ST (Z-GST).}
   \label{sgtypes}
\end{figure}

Graphene lattices and other structures, as illustrated in Fig. \ref{sgtypes}, have inspired theoretical studies on fractal Sierpinski triangle samples with different purposes such as exploring spin polarization features \cite{Perdersen2020} and the emergence of flat bands \cite{Biplab2018,Reviewflatband2018,Nandy2021}. In particular, extended eigenstates are verified in the spectrum of hexagonal ST [Fig.1 (d)], revealed as continuous bands as the lattice is threaded by magnetic fluxes \cite{Pal2012}. 
The role of fractality was already studied for diffusion, and random walks in photonic lattices \cite{XUPHOTONIC}, in the synthesis of fractal supramolecular nanostructures \cite{Wang2018}, all with interesting applications. Recently, The study paves the way for future investigations of correlated topological systems in fractal dimensions."

 One-dimensional molecular chains using Sierpinski triangles as building blocks have been reported \cite{Chain2018} using low-temperature Scanning Tunnelling Microscopy (STM). The systems were grown on Au(111) and the success of the chain formation depends on the molecular coverage and matching  between molecular size and surface lattices. Experimental methods using self-assembly\cite{JIANG2020101064,Rastgoo,JIANNATURE,Zhang2} and templating \cite{templating1,LI20151198,C7CC00566K} are being used to synthesize ST flakes and nanoribbons with transition metals and metal-organic composites. Also, 2D crystals consisting of ST units are reported by exploiting benzene-like molecules and Fe atoms on Au(111) by combining molecular design and epitaxy control \cite{2dcrystal2020}.

 Motivated by the rich experimental scenario based on different molecular constructions,  we propose theoretical models to describe such synthesized lattices. Analysis of the physical dimension is performed through the imaging process of the local density of states together with a box-counting method, revealing fractal dimensions for the studied flake systems, as expected. Results of  local electronic density of states of  simplified molecular structures are compared with spatial charge distribution reported on distinct STM images.  Here we explore electronic and transport properties of such molecular fractal chains, indicating the viability of tuning electronic states by construction, providing possible smart devices.

\section{Theoretical Model and Results}

A single-orbital tight-binding (TB) Hamiltonian is used to describe the ST systems studied, given by
\begin{gather}
    H=\sum_{i}\varepsilon_{i}c_{i}^{\dagger}c_{i}+\sum_ {\left\langle ij\right\rangle} t_{ij}c_{i}^{\dagger}c_{j}+\sum_ {\left\langle\langle ij\right\rangle\rangle} t'_{ij}c_{i}^{\dagger}c_{j}+h.c.\label{Hamito}
\end{gather}
with $\varepsilon_i$ being the on-site energy for each atom located at site $i$, $c^{\dagger}_i$ ($c_i$) is the creation (annihilation) operator of an electron on site $i$, and $t_{i,j}$  and $t'_{i,j}$ are the hopping energies for nearest and second-nearest neighboring atoms, respectively. 
%To the Chain models we have used t=0.12 eV, represented by green lines and t'=0.08*t eV, represented by black lines in Fig.\ref{stchain} and Fig.\ref{doubleuntcell}. 
Following the Green function formalism we obtain local density of states and we also investigate transport properties using the Landauer approach \cite{Data} in which the system is decoupled into three parts: central conductor and right and left leads\cite{Carbon2020,Leonor2015}. We have considered  semi-infinite ST chains as leads, matching perfectly with the central region. The central advanced (a) and retarded (r) Green functions are given as 
\begin{equation}
G^{a,r}_{c}(E)=\Big[\omega - {H_c} - \Sigma^{a,r}_L(E) - \Sigma^{a,r}_R(E)\Big]^{-1}\,\,,
\end{equation}
with $\omega = E \pm i\eta $, $\eta$ being an infinitesimal number.  ${H}_{c} $ is the Hamiltonian of the central part, and $\Sigma^{a,r}_{L,R} (E)$ correspond to left and right self-energies, given by the  related surface Green functions, from which the coupling matrices are obtained:

\begin{equation}
\Gamma^{L,R} (E)= i \Big(\Sigma_{L,R}^{r} (E) - \Sigma_{L,R}^{a}(E)\Big)\,\,.
\end{equation}

Finally, to derive the electronic conductance in ST chains, G(E)=$2e^2\mathcal{T}(E)/h$, we calculate the energy-dependent transmission given by
\begin{equation}
{\mathcal{T}(E)}=\operatorname{Tr}\left[\boldsymbol\Gamma^{L} \boldsymbol{G}^{r}_{c} \boldsymbol\Gamma^{R} \boldsymbol{G}^{a}_{c} \right]\,\,.
\end{equation}

\subsection{Conventional Flake model}

Inspired on the possibility of engineering real structures based on the geometries of the Sierpinski triangle we developed a recursive process to describe ST of different generations, based on the Green's function formalism. Our first emphasis is to investigate fractal features on the electronic properties of ST flakes. Here we show  the results of the density of states (DOS) for conventional ST flakes in terms of the order generation $l$. 
\begin{figure}[!h]
    \centering
    \includegraphics[width=8cm]{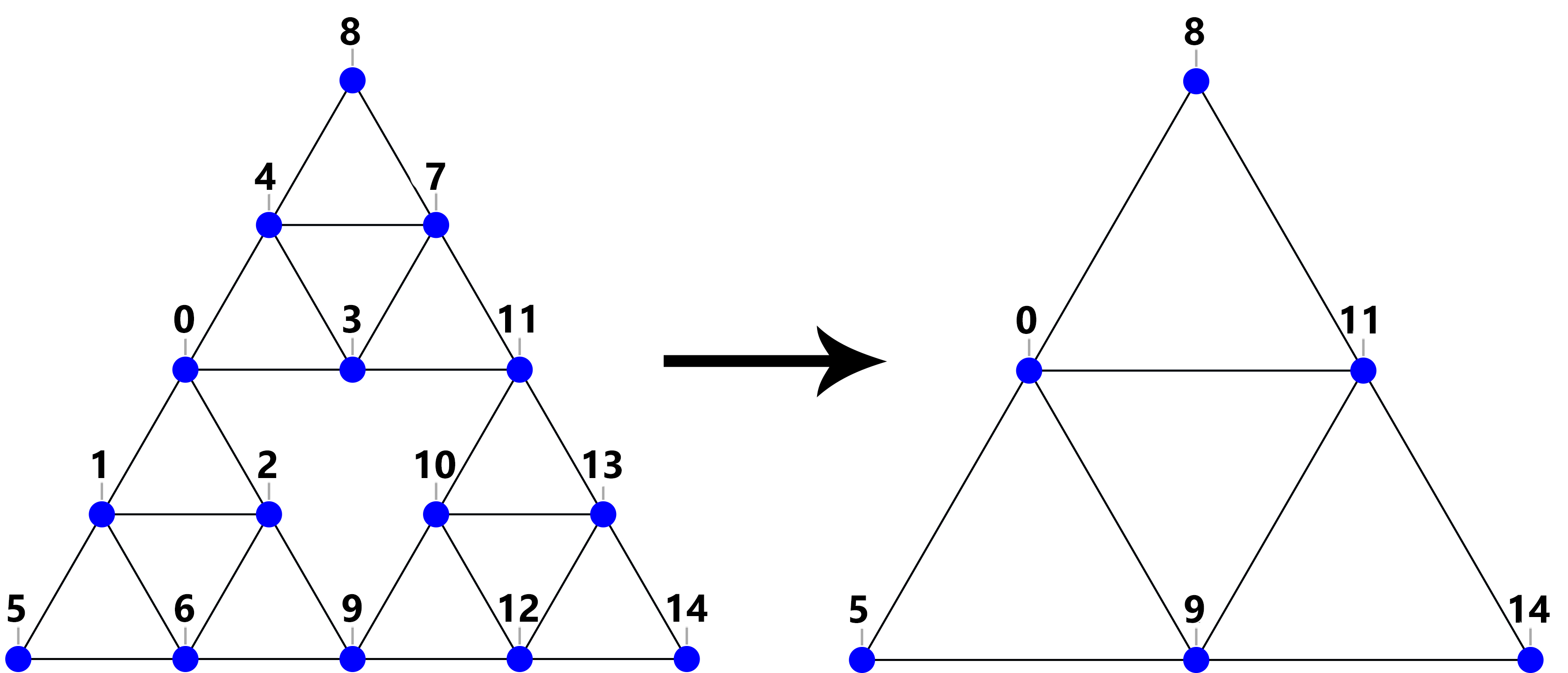}
   \caption{Illustration of the decimation method in which a G(2) ST is reduced to a G(1) ST.} 
   \label{dcmethod}
\end{figure}
\\
The  electronic properties are obtained  by  using the single first neighbor tight binding Hamiltonian, described in Eq.(1) and putting $t'=0$. The hopping energy matrices follow the corresponding connectivity of the $l$-order ST [G(l)], illustrated in Fig.\ref{dcmethod}. The DOS is given by
\\
\begin{equation}
\rho_{0}(E) = -\frac{1}{\pi} Im G_{00}(E).\label{eqDOS}
\end{equation}
\\
The Dyson equations relating the real-space Green functions, among the 15-atoms of the second ST generation, shown in Fig.2, can be written as ,
\begin{gather} 
G_{00} = g_{0} + g_{0}t[G_{10} + G_{20} + G_{30} + G_{40}]\\ \nonumber
G_{10} = g_{1}t[G_{00} +G_{20} +G_{60} +G_{50}]\\ \nonumber
G_{20} =  g_{2}t[G_{00} +G_{10} +G_{60} +G_{90}]\\ \nonumber
G_{30} = g_{3}t[G_{00} +G_{40} +G_{70} +G_{110}]\\ \nonumber
G_{40} = g_{4}t[G_{00} +G_{30} +G_{70} +G_{80}]\\ \nonumber
G_{60} = g_{6}t[G_{10} +G_{20} +G_{50} +G_{90}]\\ \nonumber
G_{70} = g_{7}t[G_{30} +G_{40} +G_{80} +G_{110}]\,\,, \nonumber
\end{gather}
where the propagator $g_i(E) = 1/(w-E_i)$ and  $t=t_{i,j}$ corresponds to the energy hopping between two first neighboring atoms i and j in the lattice, which were considered identical. After some algebraic manipulations involved in a decimation procedure\cite{Wang1995}, and considering $g_i=g_0$, for all sites i, we obtain a renormalized Dyson equation for the $G_{00}$ locator,
\begin{equation}
G_{00} = \tilde{g_{0}} + \tilde{g_{0}}\tilde{t}[G_{50} +G_{80}+G_{90} +G_{110}]\,\,,
\end{equation}
corresponding to a reduced ST generation, with the dressed propagator and hopping energy given by,
\begin{equation}
    \tilde{g_{0}}=\frac{g_{0}}{1-\frac{4g_{0}^{2}t^{2}(1+g_{0}t)}{1-3g_{0}^{2}t^{2}-2g_{0}^{3}t^{3}}}\,\,,
\end{equation}
and \nonumber
\begin{equation}
    \tilde{t}=\frac{g_{0}t^{2}(1+2g_{0}t)(1+g_{0}t)}{1-3g_{0}^{2}t^{2}-2g_{0}^{3}t^{3}}\,\,.
\end{equation}
By realizing $l$ iterative process it is possible to obtain the DOS of the $(l+1)-th$ ST generation. The results for different numbers of iterative processes (1-4) are shown in Fig.\ref{sgflakes}. 
\begin{figure}[!h]
    \centering
    \includegraphics[width=8.5cm]{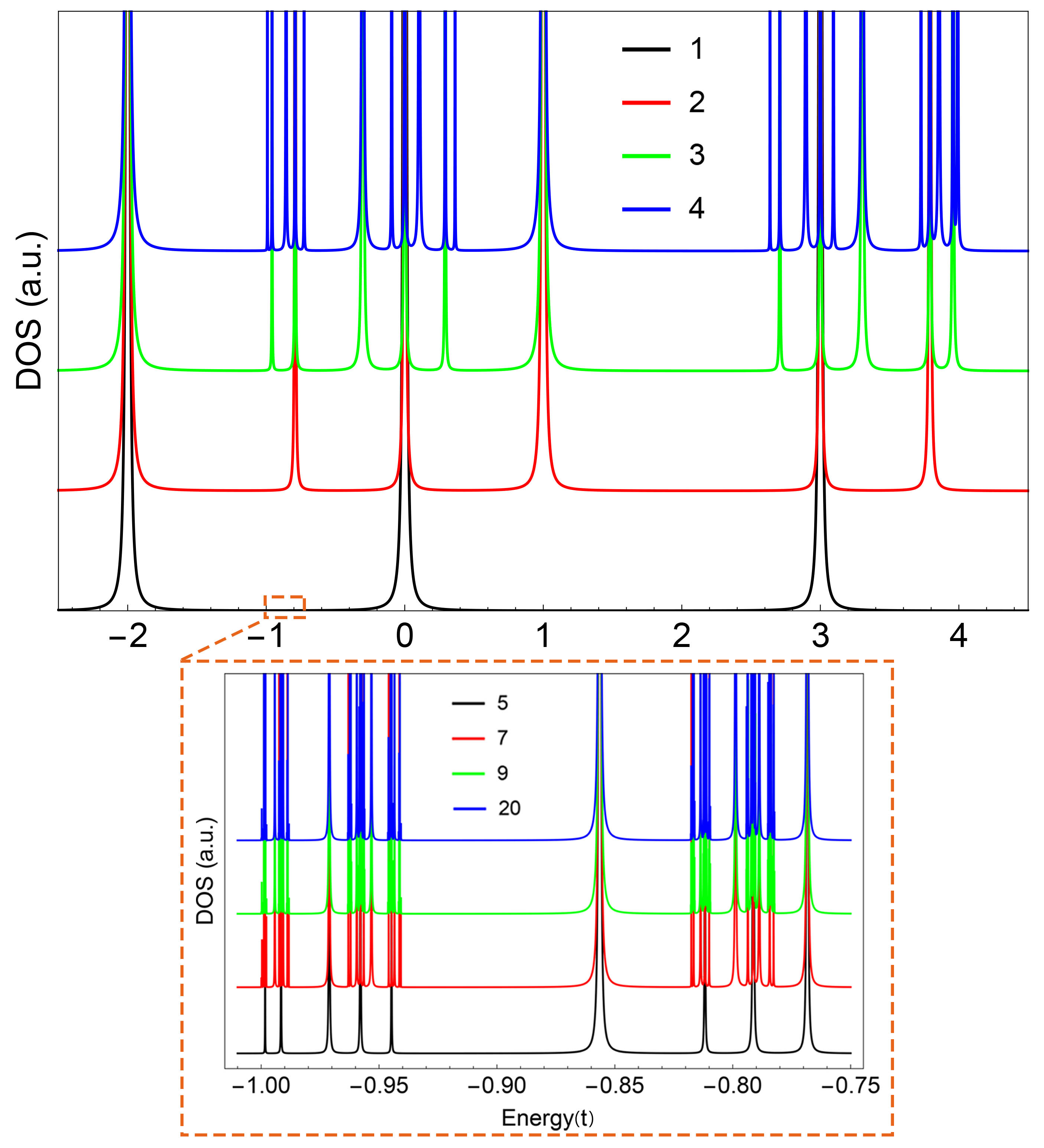}
   \caption{DOS as a function of the Fermi energy, calculated from -2.5t to 4.5t for $l$-ST varying from 1 up to 4 iterations. Zoom: DOS results within a narrower energy range(marked in the figure), for $l$=5, 7, 9 and 20 iterations, illustrating the DOS self-similarity.} 
   \label{sgflakes}
\end{figure}
 As being finite systems, the DOS are expected to exhibit a sequence of delta functions that increase as the number of atoms is increased, as depicted in Fig.\ref{sgflakes}. A self-similarity of the DOS, emerging from higher $l$-order STs is evident in the zoom (bottom panel) presenting the short energy range (-1.0t to -0.75t). Particular pinned localized states are found at the same energies, independently of the ST generation-order. More interesting to note is the fact that all states presented in a given iterative process also appear in the LDOS results of all subsequent iterative steps of higher order, in a cumulative process. We should also mention that constant energy gaps at different energy ranges are achieved as the ST flake increases, as shown in the zoom part of Fig.  \ref{sgflakes}.

\subsection{Kagomé-Honeycomb ST Flake model}

\begin{figure}[!h]
    \centering
    \includegraphics[width=8.7cm]{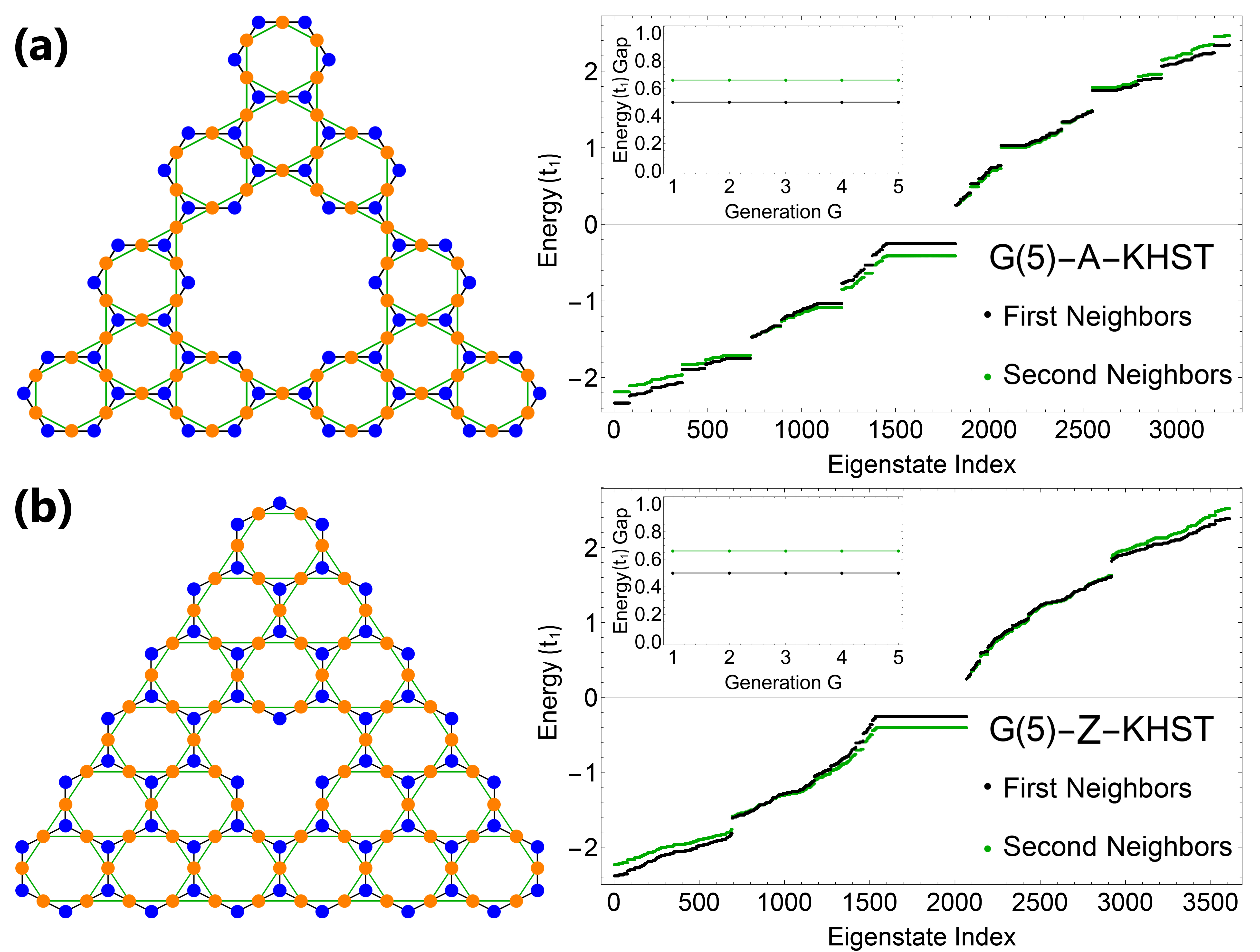}
   \caption{Schematic view of the second generation [G(2)] of an  (a) A-KHST and a (b) Z-KHST flake at left  and the eigenvalues for the 5th generation as a function of the eigenstate indexes at right. Results considering only first and up to second neighbor-hopping energies are depicted in black and green dots, respectively. Insets: gap sizes versus the ST generation.} 
   \label{zz+armkagome}
\end{figure}

Motivated by fractal properties in more realistic systems we propose the Kagomé-Honeycomb ST flake (KHST). In this case we solve the tight-binding system up to second neighbors, as illustrated in Fig.\ref{zz+armkagome}-(a) for armchair-KHST (A-KHST) and in Fig.\ref{zz+armkagome}-(b) for zigzag-KHST (Z-KHST). The number of atoms for the armchair ($N_{A}$) and zigzag ($N_{Z}$) configurations are, respectively,
\begin{gather}
N_{A}^{l}=12*3^{l}+\sum_{i=1}^{l}3^{i}\nonumber\,\,,
\\
N_{Z}^{l}=15*3^{l}+4-\sum_{i=0}^{l-1}3^{i}, \ {\rm for} \ l > 1 \ {\rm and} \ N_{Z}^{1}=49\,\,.
\end{gather}

All the calculations are made using on-site energies equal to  $\pm\epsilon= \pm0.25t_1$, with $t_1$ being the nearest-neighbor hopping, for the blue ($+\epsilon$) and orange ($-\epsilon$) sites, following reports on two-dimensional covalent organic honeycomb frameworks \cite{Chachan2016}, that are typical example of graphene-kagomé lattices. The second nearest neighbor hopping energy is chosen as \cite{nature} $t_2$=0.08$t_1$.  The LDOS is calculated as a function of energy by

\begin{equation}
    LDOS(x,y,E)=\sum_n  \abs{\Phi_n(x,y)}^2 \delta(E-E_n)\,\,,
\end{equation}
where $x$ and $y$ are the lattice position coordinates and $\Phi(x,y)$ is the corresponding electronic wave function of the $n^{th}$ state. To compute the delocalized electronic contribution around the sites, given by $\abs{\Phi_n(x,y)}^2$ we transform the delta functions into Lorentzians with $\Gamma = 0.08 t_1$. 

%(b) Eigenvalues of the Z-KHST as a function of the eingenstate index, considering only first neighbors (red dots) and up  to second-neighbor hoppings (black symbols).

For both KHST edge geometries, the presence of a band gap is verified, for the cases of first and up to second-neighbors, as shown in Fig.\ref{zz+armkagome} (right panels), revealing a semiconducting feature to such Kagomé-like ST. Otherwise, the graphene ST flake results, reveal a semiconducting and metallic characteristic for armchair and zigzag GSTs, respectively \cite{Perdersen2020} in the energy-state maps. As noticeable from Figs.\ref{zz+armkagome}(a) and (b), the gap increases by $\approx$ 0.16$t_1$ when second nearest neighbors are taken into account in the TB model, for armchair and zigzag-edged KHSTs. An interesting feature found in the electronic properties of the KHST gaskets is the reminiscent flat band observed in 2-dimensional Kagomé lattice, highlighted here by the high degeneracy order of the E=-0.25$t_1$ and -0.41$t_1$ states, for first and second-neighbor models, respectively, at the gap threshold. The number of such localized degenerate states is expected to increase as the generation order of the gasket is increased, although the size of the gaps do not dependent on the generation order, as depicted in the insets of Fig.\ref{zz+armkagome}, for armchair and zigzag configurations. Compared with the gap dependence for GST flakes \cite{Perdersen2020}, for which the gap size saturates for $l$ between 3 and 4, we must comment that the  number of atoms in the KHST flakes are considerable superior at the same generation order, what may justify the constant gap size achieved already in the first generation.

\begin{figure}[!h]
    \centering
    \includegraphics[width=8.7cm]{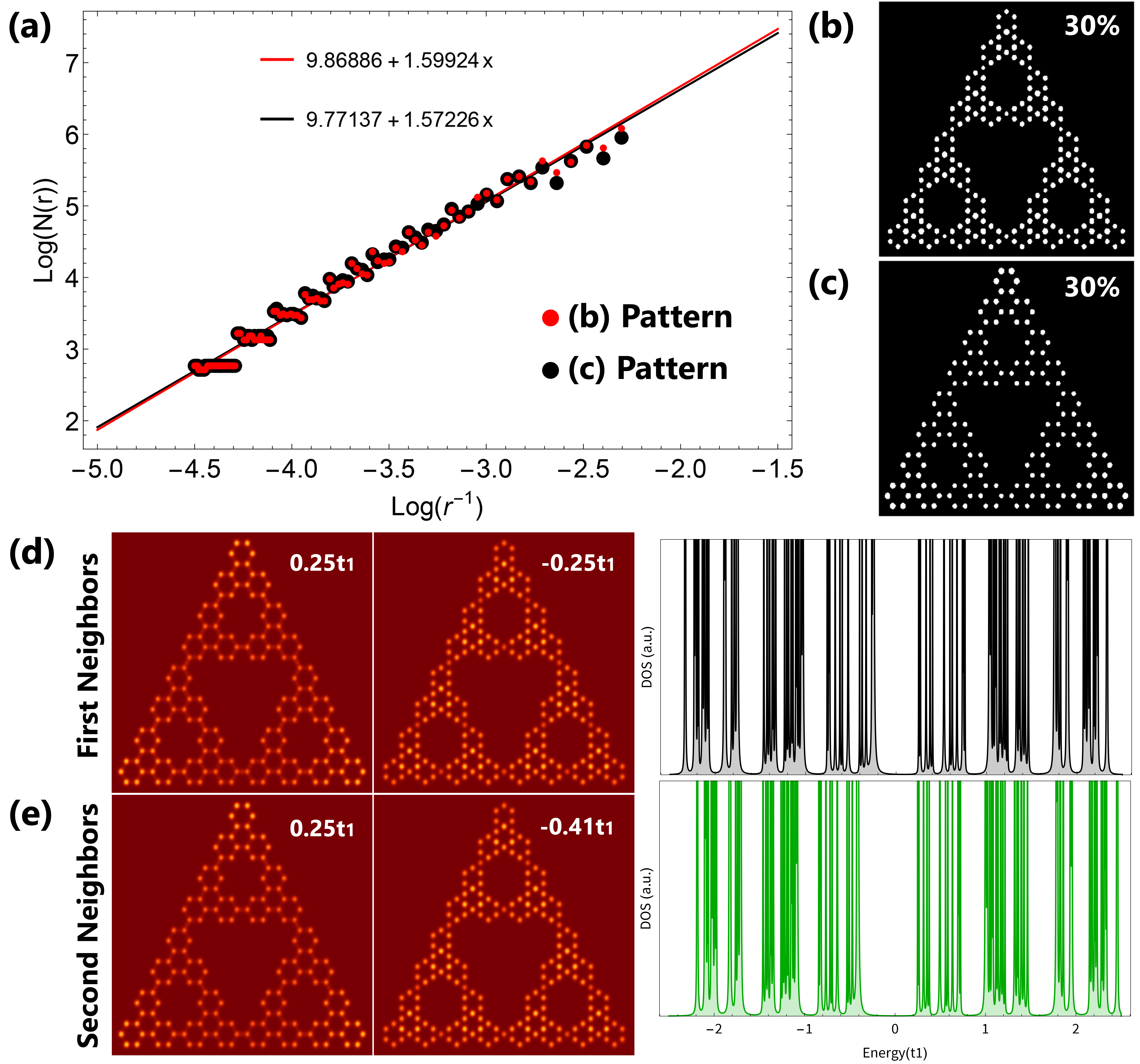}
   \caption{(a) Box-counting results of the binarized  images (b) and (c) at  $E=\pm 0.25 t_1$, represented by red and black colors, respectively. LDOS of a G(3)-A-KHST for (d) first neighbors at $E=\pm0.25t_1$ and (e) second neighbors at $E=0.25t_1$ and $E=-0.41t_1$ and corresponding  DOS as function of the energy.} 
   \label{HAUSDORFF}
\end{figure}

Due to the difference in the geometry and site numbers in the basic units forming the KHSTs in comparison to the conventional ST, another fractal dimension definition is used \cite{FOROUTANPOUR1999195}. Here we adopt the Minkowski–Bouligand dimension used for monofractals, given by  

\begin{equation}
    D=\lim_{r \to 0} \frac{Log(N(r))}{Log(r^{-1})}\,\,,
\end{equation}
with $N(r)$ and $r$ being the number of squares covering the full system, and the square size, respectively. These parameters are used in the box-counting method \cite{forum} to perform the dimension calculation. The dimension of each system is determined by the line slope coefficient of $Log(N(r))$ vs $Log(r^{-1})$ graph presented in Fig.\ref{HAUSDORFF}(a).

Starting from the atomic spatial localization in the A-KHST flake (see Fig.\ref{sgtypes}(e)), the sizes  (in pixels) of each square found to better adjust the slope are 10 up to 90 pixels. In both cases we take the LDOS image (360 x 360 pixels) and binarize it in a threshold of  $30\%$, to represent the most accurate electronic distribution. The details of such numerical process is described in Refs. \cite{nature,FOROUTANPOUR1999195}. Perfect Kagomé nanoribbons and Kagomé 2D-lattices, exhibit symmetric band spectra, except for the presence of a flat band located exactly at $\epsilon$ or -$\epsilon$, depending on the hopping energy signal. For this reason, in this work we choose then the values of $E=\pm \epsilon$ to calculate the maximum contribution in LDOS indicated by its brightest spots, which gives two different patterns, as can be seen in the binarized image shown in Fig.\ref{HAUSDORFF}-(b,c).

%(a) and (b)/(c) and (d) LDOS for first/second neighbors of A-KHST Flake, with the according DOS in red/black, respectively. (e) and (f) Binarized image of both LDOS (-0.25/$t_1$) in (a) and (c) with a threshold of 30$\%$. (g) Slope and points of each case are represented by the same DOS colors. The Hausdorff dimension is obtained by the slope of both lines D$\approx$1.58.

In this system we have two energy values that the maximum contribution in LDOS is spread through the flake, given origin to the patterns (b) and (c) in Fig.5. Taking them both in the analysis, the box counting method yields a fractal dimension close to the conventional Sierpinski triangle gasket, $\approx 1.585$, as can be verified by the slope values shown in Fig.\ref{HAUSDORFF}(a), for both situations ($1^{st}$ and up to $2^{nd}$ neighbor hopping). This occurs because the brightest spots in the LDOS maps are near from the geometrical points of a A-GST flake, that are also fractal. Otherwise, for energies where the bright spots (high LDOS values) do not reveal the real geometry of the flake, the fractal dimension varies between D=1.30-1.80, as verified by Kempkes \textit{et al.}\cite{nature} in hexagonal flakes, varying between low and high contributions of each site for the LDOS, respectively. The self-similarity observed in others ST flakes \cite{nature,Perdersen2020} is also evidenced here for both configurations of the KHSTs (armchair and zigzag), independently on the order of the neighboring hopping taken into account.

A further interesting feature to explore in zigzag and armchair KHST flakes, described by the first and second neighbors approximation, is the asymmetric distribution of the electronic probability distribution at $E_1$=-0.25$t_1$ and $E_2$=0.25$t_1$ states.
Considering the first generation G(1), we found that while the $E_1$ state is formed by four-fold degenerate eigenstates, the $E_2$ state is non-degenerated. These states are spread differently among the orange and blue sites of the flake [see Fig.  \ref{zz+armkagome}] , giving raise to the LDOS exhibited in Fig.\ref{HAUSDORFF}(d) and (e), respectively. The same behavior is evidenced for higher orders of the KHSTs, considering both first and up to second-neighbor hopping models. We would like to emphasize  that the inclusion of second neighbor hopping in our tight binding description did not promote significant changes on the main physical responses explored for the Kagomé-like ST flakes.

\subsection{ST Mirrored Chains}
\begin{figure}[!h]
    \centering
    \includegraphics[width=8.7cm]{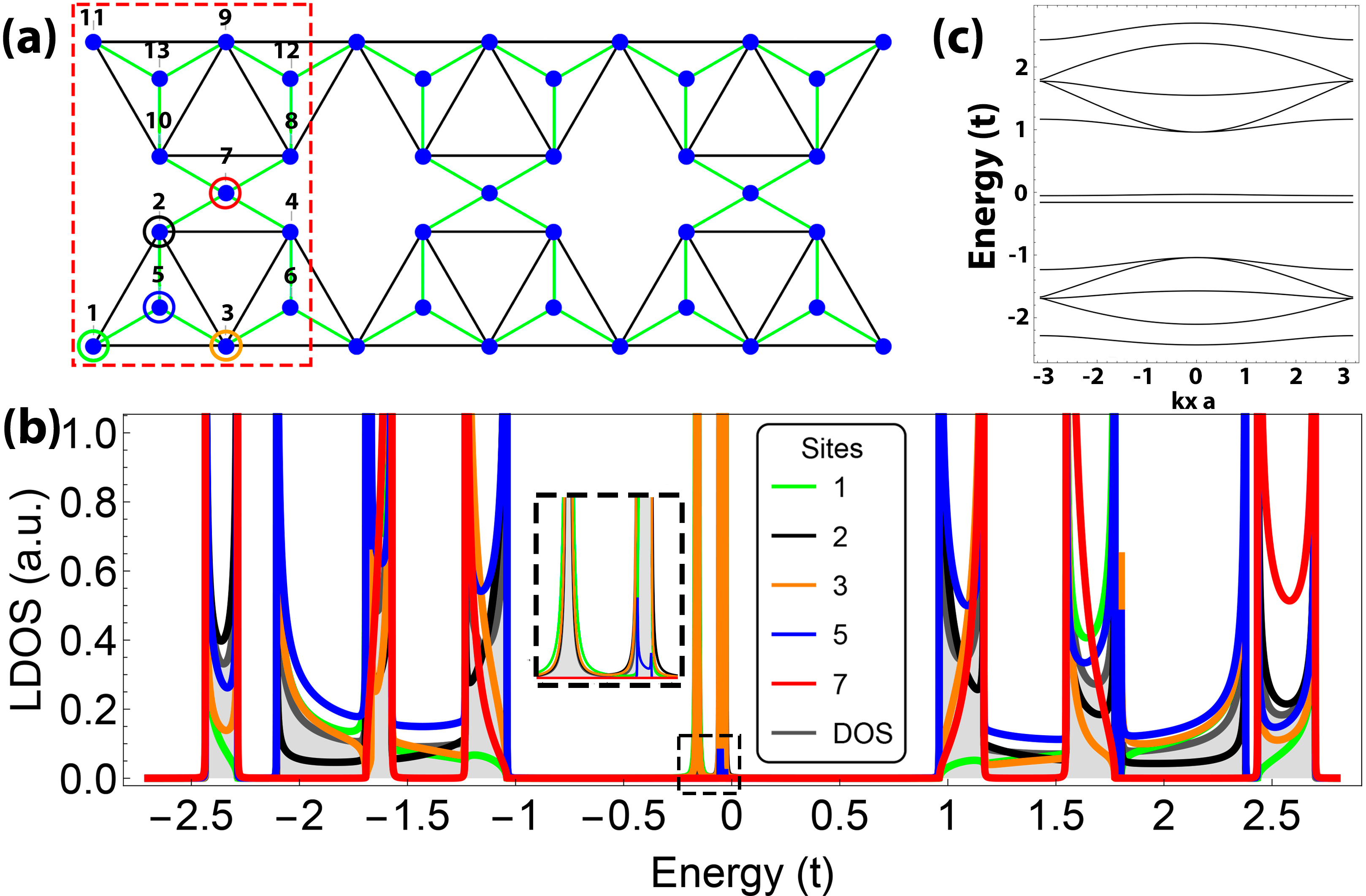}
   \caption{(a) Effective G(1)-ST Chain. The green and black lines are related to hopping t and t', respectively. The system extends infinitely along the x-direction. (b) LDOS in function of the energy at sites 1,2,3,5 and 7 identified and colored circled inside the red dashed unit-cell of the ST-Chain, defined in part (a). The total DOS is also shown with pink shadow regions.  (c) Band structure of the ST chain for $\epsilon$=0.} 
   \label{stchain}
\end{figure}

Following the experimental realization of 1-D molecular chains \cite{Chain2018}, with ST as building blocks grown on Au(111), we explore electronic properties of a similar quasi-1D chain as depicted in Fig \ref{stchain}(a). The system was idealized based on the coupling of two hexagonal STs spatially inverted (up and down), with mirror symmetry. Green and black lines connect nearest and second-nearest neighboring atoms, respectively, through the hopping terms $t$ and $t'$ in the tight binding Hamiltonian given in Eq.(1). Differently from the long-range ordered grown structures where Co atoms are present in intercalated benzene lattices\cite{Chain2018}, our simple model involves exclusively carbon atoms, denoted by blue dots. The electronic properties are calculated using on site energy $\epsilon=0$ and following the relation t'=0.08t  for the hopping parameters as recently for HSTs \cite{nature}.

The LDOS shown in Fig.\ref{stchain}-(b) is calculated via Eq.\ref{eqDOS}, where $G_{00}$ is obtained using similar real-space decimation methods, properly constructed to infinite periodic systems \cite{Latge2008,Latge2015,Latge2021}. At energies close to the Fermi level, the electronic group velocity is near zero, giving origin to almost two dispersionless  bands as can be seen in the corresponding electronic structure shown in Fig.\ref{stchain}-(c). These flat bands appear as highly peaked density of states, and are highlighted in the inset in Fig.\ref{stchain}-(b). The different curves correspond to the assigned sites marked with the same color in the unit cell displayed in part (a). As seen in Fig.\ref{stchain}-(b), at E=-0.16$t$, the main contribution becomes from sites 1, 2 and 3, and from  the symmetric upper sites 11, 10 and 9.

\begin{figure}[!h]
    \centering
    \includegraphics[width=8.6cm]{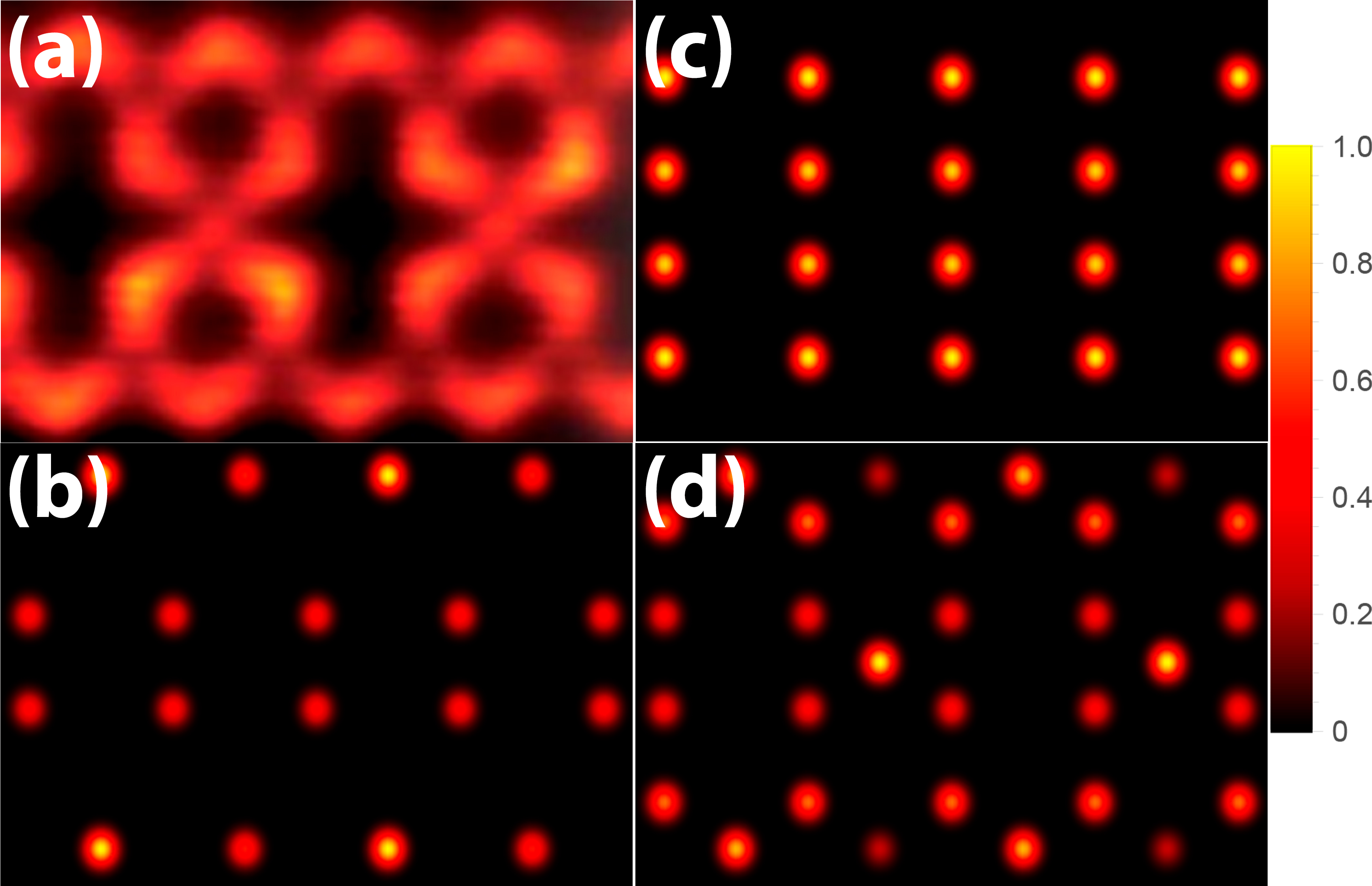}
   \caption{LDOS contour plots of a G(1)-ST Chain. a) STM image adapted from the experimental data reported in Ref.\cite{Chain2018}. Our theoretical LDOS results for (b) E=-0.16$t$, (c) 0.96$t$, and (d) 1.65$t$.} 
   \label{ldosst}
\end{figure}

A better visualization of the spatial electronic distribution through the system is displayed in Fig.\ref{ldosst} for distinct energies. While part (a) refers to a STM image adapted from ref. \cite{Chain2018}, Figs.\ref{ldosst} (b), (c), and (d), are the LDOS theoretical results  at $E=-0.16t$ (flat band), $0.96t$ and $1.65t$. In the STM image a high charge distribution appears as bright spots at the relative atomic sites. Correspondingly, in our results the color maps refer to the LDOS intensity. The normalized LDOS results shown in Fig.\ref{ldosst} (b) reveal that at one of the flat bands [E=-0.16t] not all the atomic sites of the lattice are populated, in according with the previously discussion. This last state appears again in the Dephased ST Chain, and in special, it will not contribute to the transport properties of these systems, as we will see in the next session. 

Moreover, while at $E=0.96t$, only sites 2 and 5, and their equivalents in the unit cell, contribute to the state  [see Fig.\ref{stchain}-(c)], revealing a  restricted charge distribution, for the state $E=1.65t$, Fig.\ref{stchain}-(d), almost all the lattice is visited generating bright spots in the geometrical net. The later LDOS pattern strongly resembles the cited STM image [Fig.\ref{ldosst}-(a)] indicating that our simple model can be an important tool of modelling such organic molecular chain.

\begin{figure}[!h]
    \centering
    \includegraphics[width=8 cm]{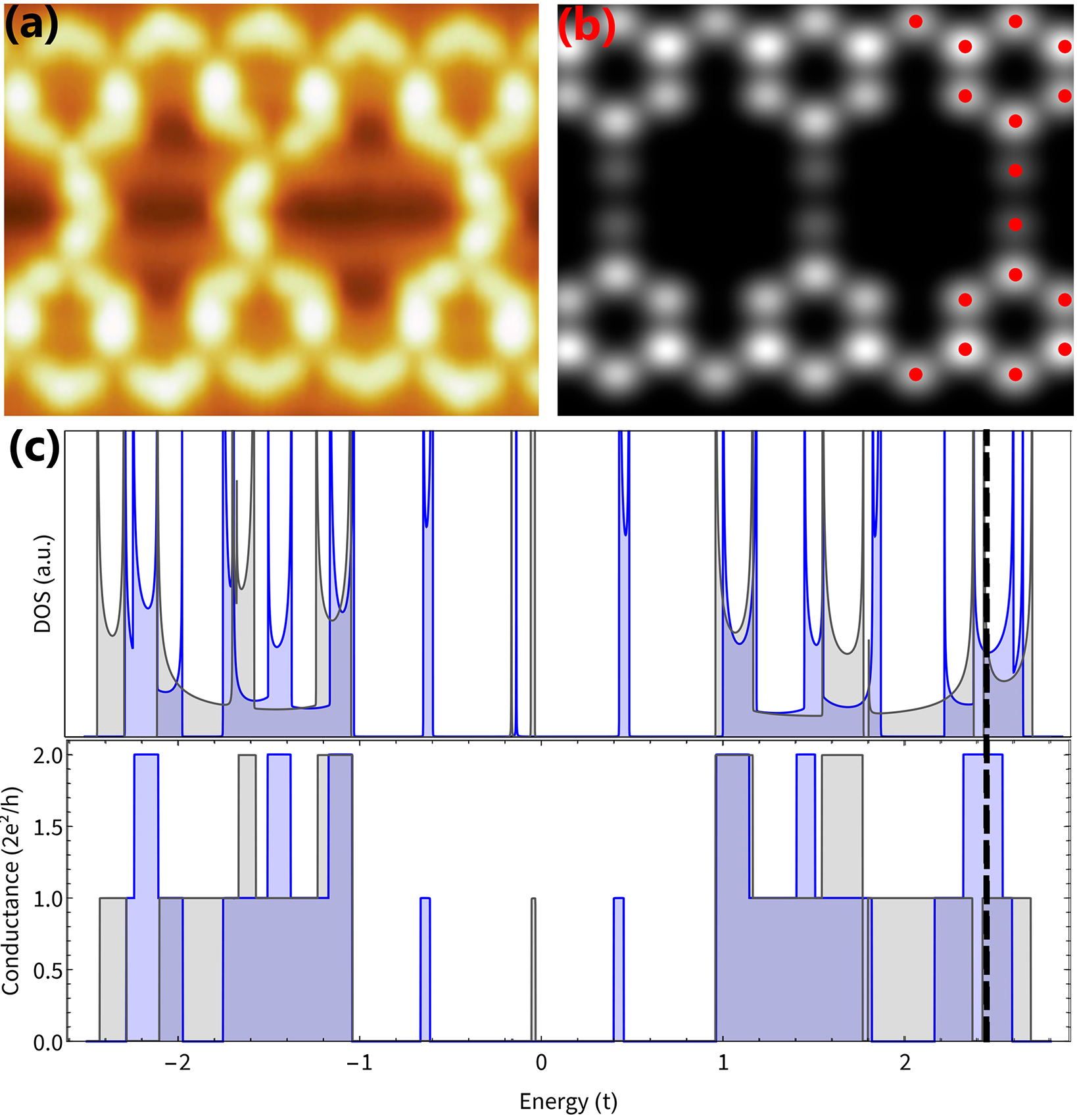}
   \caption{(a) STM images of the MSTC-2 [Adapted from \cite{Liu2017}]. (b) LDOS of the MSTC for a Fermi energy equal to 2.4t. The  nanoribbon unit cell is highlighted with red dots. (c) DOS and conductance results of the MSTC and MSTC-2 in blue and gray, respectively.} 
   \label{stchainalternative}
\end{figure}

Another experimental alternative of generating 1D mirrored chains made of Sierpiński triangles was proposed in Ref.\cite{Liu2017}. We call it as  mirrored chain structure-type 2 (MSTC-2). In our model up and down triangles are now fully preserved resembling the STM images shown in Fig. \ref{stchainalternative}(a).    The LDOS at a particular energy ($E=2.4t$) is depicted in part (b) where the region enclosing the nanoribbon unit cell is marked with red circles. The results for DOS and conductance are shown in Fig.\ref{stchainalternative} (c), highlighted with shadowed blue curves and compared with the results found for the previous molecular chain [shaded gray curves]. It is noticeable the emergence of two narrowed energy states near the Fermi Level for the MSTC-2, this is a characteristic property of double ST Chains that will be explored in next section. Also, the flat-like states in both chain configurations are preserved in the central gap in DOS, although they are suppressed in the electronic conductance response of each molecular designs due to the high localization features of the flat band.

 %-----------------------------------

 \subsection{Dephased ST Chain}
\begin{figure}[!h]
    \centering
    \includegraphics[width=8.5cm]{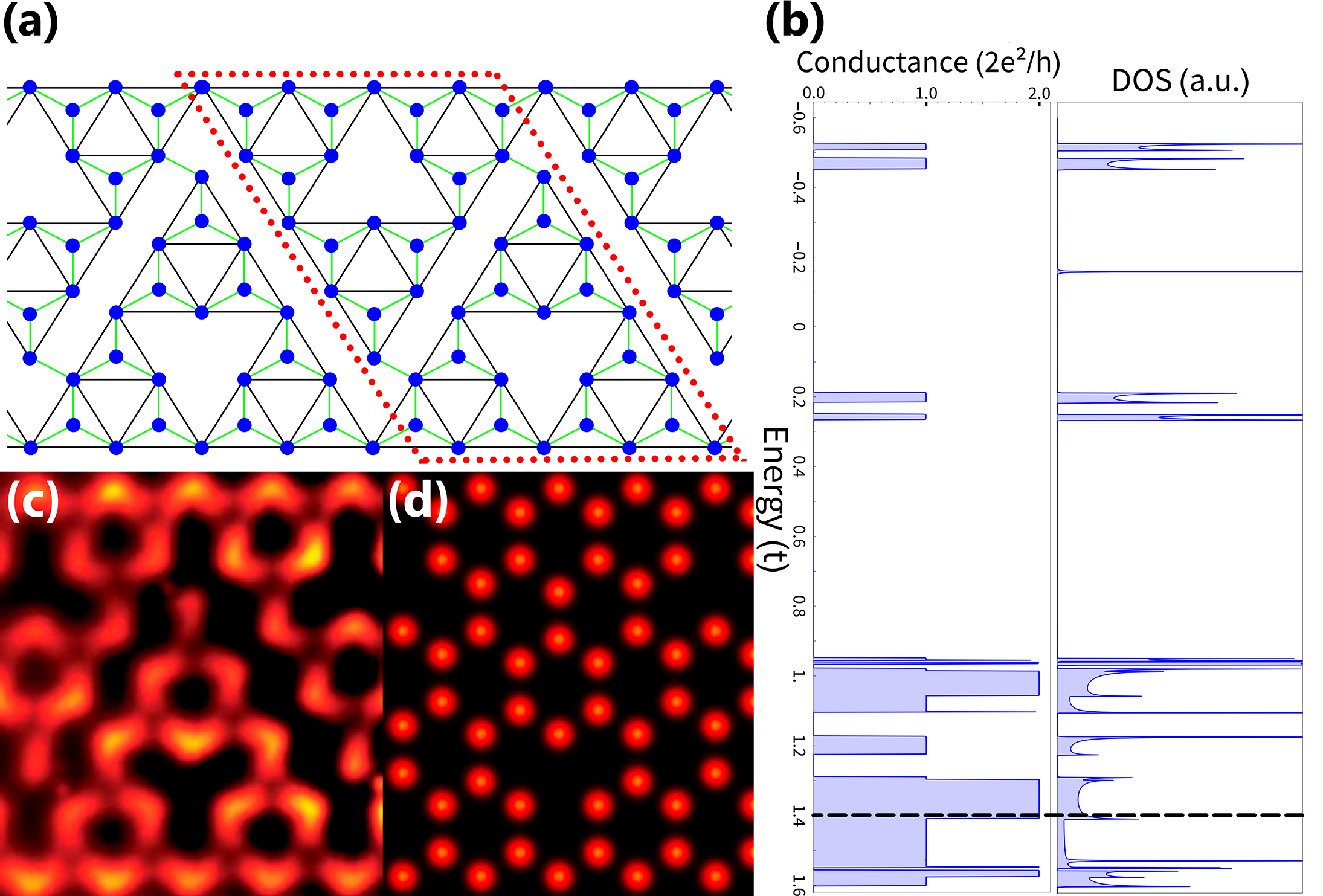}
   \caption{(a) Effective G(2)-dephased ST chain, (b) Conductance and DOS, (c) STM image [adapted from Ref. 15], and (d) LDOS at E=1.4t.} 
   \label{g1g2g3comparison}
\end{figure}

 Following the experimental molecular structures presented in Ref. \cite{Chain2018}, we address now another proposal for ST chain, as illustrated in Fig. \ref{g1g2g3comparison}-(a). Differently from the previous discussed molecular chains, the new structure are composed by dephased pairs of HSTs [G(2)], connected at particular lateral lattice sites, simulating the packing mode of STs produced by a combination of a Co atom and three BPyB molecules\cite{Chain2018}. The unit cells of the dephased molecular chain is marked  with dashed red lines. It is interesting to note that our theoretical result for the LDOS at $E=1.4t$, as shown in Fig.\ref{g1g2g3comparison}-(d),  reproduces quite well the STM image presented in Fig.\ref{g1g2g3comparison}-(c) for the grown molecular nanostructured, reported in Ref.\cite{Chain2018}. 

 The electronic properties of the G(2)-dephased ST is shown in Fig. \ref{g1g2g3comparison}-(b) via the density of states and conductance results. Due to the high electronic localization, the DOS peak at E=-0.16t does not contribute to the electronic conductance in G(2), what happens also for dephased chains of higher ST orders (not shown here). 
  %%%%%%%%%%%%%%%%%%%%%%%%%%%%%%%%%%%%%%%%%%%%%%%%%%%
 
 %It is interesting to note that quite similar local density of state patterns  are derived by following the simple adopted tight binding hamiltonian and controlling the electronic features given in spatial maps by handing with energy hopping parameters.

\begin{figure}[!h]
   \centering
   \includegraphics[width=8.7cm]{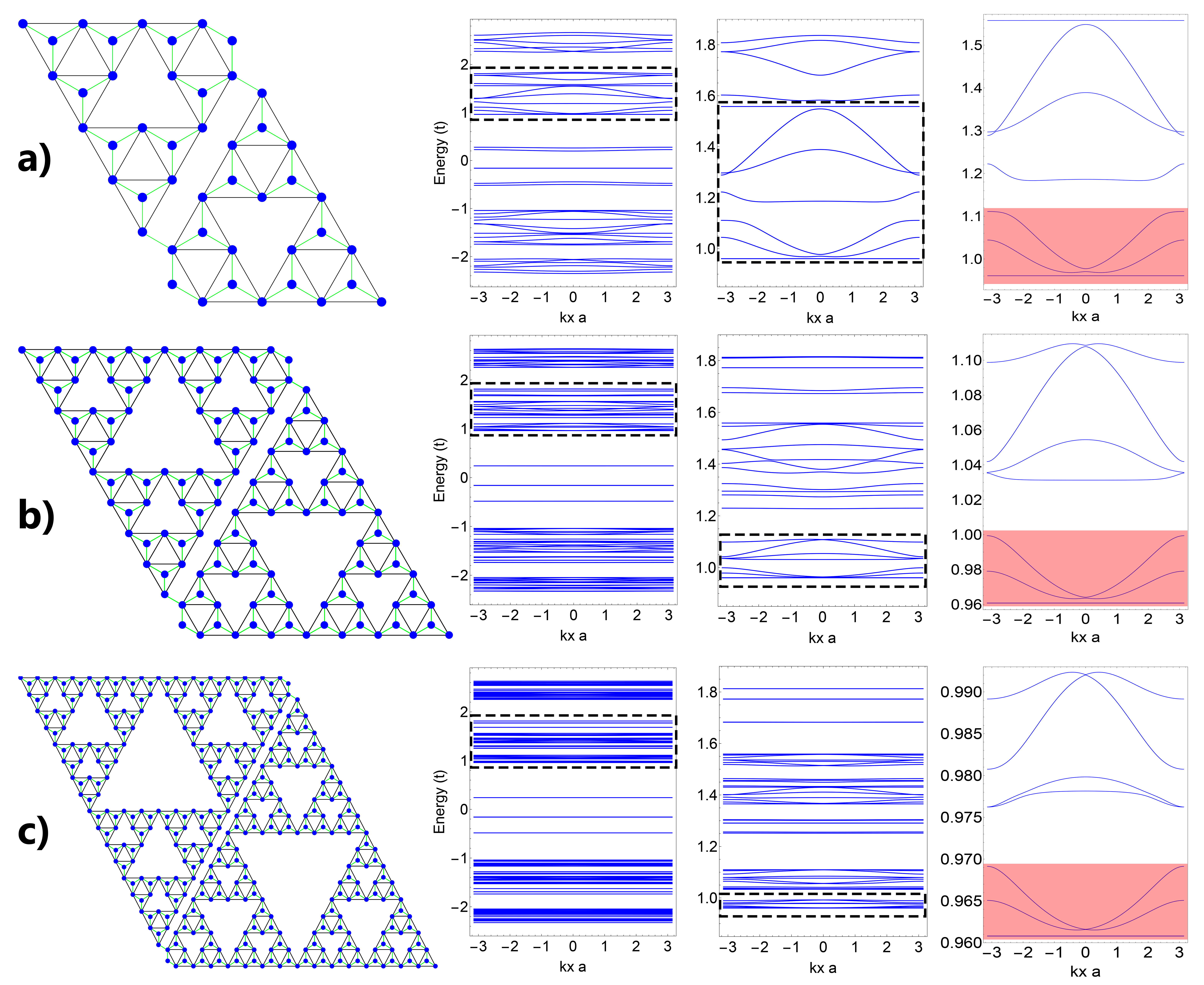}
   \caption{(a-c) Unit cells of G(2-4) Dephased ST Chains and corresponding band structures, followed by its zooms, from left to right.}    \label{g4g3g2}
\end{figure}

 To investigate fractal properties on such nanostructured systems we concentrate now in the electronic band structures. We present in Fig. \ref{g4g3g2} (a-c) unit cells of three generations of the proposed molecular chain discussed in Fig.\ref{g1g2g3comparison}, G(2), G(3) and G(4), and the corresponding band structure of the 1D chains. The dashed regions marked in the band structures presented in Fig.\ref{g4g3g2}(a), related to the G(2) lattice, was enlarged in the subsequent panels at right, revealing a single flat band and other two bands presenting a small energy dispersion, at the lower energy range, highlighted by the red shaded area. These later are also noted in the DOS and conductance results (not shown).  To analyze the role played by the generation order on the used HST blocks forming the chains, we show in Fig.\ref{g4g3g2} (b) and (c), for comparison, the energy bands of the three G(3) and G(4) dephased ST chains. The auto similarity of these systems can be evidenced looking firstly, at the same energy region of the three generation proposed, defined by the dashed energy range. Comparing the results, we notice in the last right panels that the common pattern of a single flat band (happening at 0.96t in the three cases), followed by two dispersive bands [ending at 1.1t (a), 1.0t (b) and 0.97t (c), respectively] is slightly preserved going from G(2) to G(4).  As expected, for increasing ST generations, the repeated band structure pattern are found at smaller energy ranges.
 
 Moreover, the results for the energy bands reveal an increasing number of 1D-like van Hove singularities passing from G(2) to G(4) dephased chains. Similarly as reported before \cite{Biplab2018}, the number of flat bands increases significantly with the ST order used to build the chains. Our findings suggest that properly tuning the tight binding parameters and playing with different connections between and inside the unit cells, better modeling of such nanostructured systems may be achieved.  In particular,  considering different on-site energies for Co atoms and BPyB molecules in our model (for instance taking $E_C=-E_{Co}$), simplified here as carbon atoms, only promotes a shift in the double flat states near the fermi level. We would like also to draw attention to the fact that temperature and different types  of lattice imperfections, such as edge roughness, may limit the experimental determination of the gap hierarchy gaps with decreasing size upon increasing the order generation of the structure, mainly due to a loss of quantum coherence and the consequent  inhomogeneous broadening of electronic levels. As previously addressed in Ref. \cite{macia} these factors can introduce a critical size of the system, above which finer details of the fractal electronic spectrum may be difficult to the revealed.

%--------------------------------------
\section{Conclusion}

We have explored fractal features in the electronic properties of ST flakes and molecular chains of different geometry details. The results found for the fractal dimension of the Kagomé-like ST flakes follow recent reports on graphene-like STs. The result is corroborated by the Kagomé ST LDOS at particular energy states that exhibit typical graphene  ST spatial configurations. The self similarity of the energy states are found comparing different ST generation orders and also amplifying  the energy ranges investigated, for both flakes and quasi 1D systems. In particular, the results for the local density of states of the theoretical  molecular chains proposed here exhibit quite similar spatial charge distribution  as experimental STM reports. The analysis of transport response of such quasi 1D molecular chains reveals localized states that do not contribute to the electronic transport. The study  can be used as guide to propose a variety of architecture in the synthesis of real molecular chains.

 Further aspects of the theoretical framework adopted were explored such as changing the inter and intra unit cell connections of the proposed molecular chains. Some of the schemes induce degeneracy break of particular electronic states, favoring the disrupt of flat bands. Although a primary analysis on more realistic on-site energies, simulating metallic atoms in the ST chains that appear in the experiments, has not indicated great changes, more sophisticated theoretical framework must be used for a fully study of tuning the energy channel positions in the transport features of the chains. Probably, the inclusion  of electron-electron interaction in the model will bring important light into the electronic occupation in conductive  states, in special, in the flat states presented.

%-------------------------------------
\section*{Acknowledgments}
%---------------------------------------------

This work was financially supported by the Brazilian Agencies CAPES, CNPq and by FAPERJ under the grant E-26/202.567/2019, and the INCT de Nanomateriais de Carbono.

\bibliography{refs}

\begin{thebibliography}{37}%
\makeatletter
\providecommand \@ifxundefined [1]{%
 \@ifx{#1\undefined}
}%
\providecommand \@ifnum [1]{%
 \ifnum #1\expandafter \@firstoftwo
 \else \expandafter \@secondoftwo
 \fi
}%
\providecommand \@ifx [1]{%
 \ifx #1\expandafter \@firstoftwo
 \else \expandafter \@secondoftwo
 \fi
}%
\providecommand \natexlab [1]{#1}%
\providecommand \enquote  [1]{``#1''}%
\providecommand \bibnamefont  [1]{#1}%
\providecommand \bibfnamefont [1]{#1}%
\providecommand \citenamefont [1]{#1}%
\providecommand \href@noop [0]{\@secondoftwo}%
\providecommand \href [0]{\begingroup \@sanitize@url \@href}%
\providecommand \@href[1]{\@@startlink{#1}\@@href}%
\providecommand \@@href[1]{\endgroup#1\@@endlink}%
\providecommand \@sanitize@url [0]{\catcode `\\12\catcode `\$12\catcode
  `\&12\catcode `\#12\catcode `\^12\catcode `\_12\catcode `\%12\relax}%
\providecommand \@@startlink[1]{}%
\providecommand \@@endlink[0]{}%
\providecommand \url  [0]{\begingroup\@sanitize@url \@url }%
\providecommand \@url [1]{\endgroup\@href {#1}{\urlprefix }}%
\providecommand \urlprefix  [0]{URL }%
\providecommand \Eprint [0]{\href }%
\providecommand \doibase [0]{http://dx.doi.org/}%
\providecommand \selectlanguage [0]{\@gobble}%
\providecommand \bibinfo  [0]{\@secondoftwo}%
\providecommand \bibfield  [0]{\@secondoftwo}%
\providecommand \translation [1]{[#1]}%
\providecommand \BibitemOpen [0]{}%
\providecommand \bibitemStop [0]{}%
\providecommand \bibitemNoStop [0]{.\EOS\space}%
\providecommand \EOS [0]{\spacefactor3000\relax}%
\providecommand \BibitemShut  [1]{\csname bibitem#1\endcsname}%
\let\auto@bib@innerbib\@empty
%</preamble>
\bibitem [{\citenamefont {Chandel}\ \emph {et~al.}(2020)\citenamefont
  {Chandel}, \citenamefont {Wang},\ and\ \citenamefont {Talha}}]{Review2020}%
  \BibitemOpen
  \bibfield  {author} {\bibinfo {author} {\bibfnamefont {V.~S.}\ \bibnamefont
  {Chandel}}, \bibinfo {author} {\bibfnamefont {G.}~\bibnamefont {Wang}}, \
  and\ \bibinfo {author} {\bibfnamefont {M.}~\bibnamefont {Talha}},\ }\href
  {\doibase doi:10.1515/ntrev-2020-0020} {\bibfield  {journal} {\bibinfo
  {journal} {Nanotechnology Reviews}\ }\textbf {\bibinfo {volume} {9}},\
  \bibinfo {pages} {230} (\bibinfo {year} {2020})}\BibitemShut {NoStop}%
\bibitem [{\citenamefont {Li}\ \emph {et~al.}(2022)\citenamefont {Li},
  \citenamefont {Han}, \citenamefont {Qin}, \citenamefont {Xiong},
  \citenamefont {Huang}, \citenamefont {Wang}, \citenamefont {Ding},
  \citenamefont {Hu}, \citenamefont {Xu},\ and\ \citenamefont
  {Zhu}}]{Kagome2022}%
  \BibitemOpen
  \bibfield  {author} {\bibinfo {author} {\bibfnamefont {X.}~\bibnamefont
  {Li}}, \bibinfo {author} {\bibfnamefont {D.}~\bibnamefont {Han}}, \bibinfo
  {author} {\bibfnamefont {T.}~\bibnamefont {Qin}}, \bibinfo {author}
  {\bibfnamefont {J.}~\bibnamefont {Xiong}}, \bibinfo {author} {\bibfnamefont
  {J.}~\bibnamefont {Huang}}, \bibinfo {author} {\bibfnamefont
  {T.}~\bibnamefont {Wang}}, \bibinfo {author} {\bibfnamefont {H.}~\bibnamefont
  {Ding}}, \bibinfo {author} {\bibfnamefont {J.}~\bibnamefont {Hu}}, \bibinfo
  {author} {\bibfnamefont {Q.}~\bibnamefont {Xu}}, \ and\ \bibinfo {author}
  {\bibfnamefont {J.}~\bibnamefont {Zhu}},\ }\href@noop {} {\bibfield
  {journal} {\bibinfo  {journal} {Nanoscale}\ }\textbf {\bibinfo {volume}
  {14}},\ \bibinfo {pages} {6239} (\bibinfo {year} {2022})}\BibitemShut
  {NoStop}%
\bibitem [{\citenamefont {Lisi}\ \emph {et~al.}(2021)\citenamefont {Lisi},
  \citenamefont {Lu}, \citenamefont {Benschop}, \citenamefont {de~Jong},
  \citenamefont {Stepanov}, \citenamefont {Duran}, \citenamefont {Margot},
  \citenamefont {Cucchi}, \citenamefont {Cappelli}, \citenamefont {Hunter},
  \citenamefont {Tamai}, \citenamefont {Kandyba}, \citenamefont {Giampietri},
  \citenamefont {Barinov}, \citenamefont {Jobst}, \citenamefont {Stalman},
  \citenamefont {Leeuwenhoek}, \citenamefont {Watanabe}, \citenamefont
  {Taniguchi}, \citenamefont {Rademaker}, \citenamefont {van~der Molen},
  \citenamefont {Allan}, \citenamefont {Efetov},\ and\ \citenamefont
  {Baumberger}}]{Twisted2021}%
  \BibitemOpen
  \bibfield  {author} {\bibinfo {author} {\bibfnamefont {S.}~\bibnamefont
  {Lisi}}, \bibinfo {author} {\bibfnamefont {X.}~\bibnamefont {Lu}}, \bibinfo
  {author} {\bibfnamefont {T.}~\bibnamefont {Benschop}}, \bibinfo {author}
  {\bibfnamefont {T.~A.}\ \bibnamefont {de~Jong}}, \bibinfo {author}
  {\bibfnamefont {P.}~\bibnamefont {Stepanov}}, \bibinfo {author}
  {\bibfnamefont {J.~R.}\ \bibnamefont {Duran}}, \bibinfo {author}
  {\bibfnamefont {F.}~\bibnamefont {Margot}}, \bibinfo {author} {\bibfnamefont
  {I.}~\bibnamefont {Cucchi}}, \bibinfo {author} {\bibfnamefont
  {E.}~\bibnamefont {Cappelli}}, \bibinfo {author} {\bibfnamefont
  {A.}~\bibnamefont {Hunter}}, \bibinfo {author} {\bibfnamefont
  {A.}~\bibnamefont {Tamai}}, \bibinfo {author} {\bibfnamefont
  {V.}~\bibnamefont {Kandyba}}, \bibinfo {author} {\bibfnamefont
  {A.}~\bibnamefont {Giampietri}}, \bibinfo {author} {\bibfnamefont
  {A.}~\bibnamefont {Barinov}}, \bibinfo {author} {\bibfnamefont
  {J.}~\bibnamefont {Jobst}}, \bibinfo {author} {\bibfnamefont
  {V.}~\bibnamefont {Stalman}}, \bibinfo {author} {\bibfnamefont
  {M.}~\bibnamefont {Leeuwenhoek}}, \bibinfo {author} {\bibfnamefont
  {K.}~\bibnamefont {Watanabe}}, \bibinfo {author} {\bibfnamefont
  {T.}~\bibnamefont {Taniguchi}}, \bibinfo {author} {\bibfnamefont
  {L.}~\bibnamefont {Rademaker}}, \bibinfo {author} {\bibfnamefont {S.~J.}\
  \bibnamefont {van~der Molen}}, \bibinfo {author} {\bibfnamefont {M.~P.}\
  \bibnamefont {Allan}}, \bibinfo {author} {\bibfnamefont {D.~K.}\ \bibnamefont
  {Efetov}}, \ and\ \bibinfo {author} {\bibfnamefont {F.}~\bibnamefont
  {Baumberger}},\ }\href {\doibase 10.1038/s41567-020-01041-x} {\bibfield
  {journal} {\bibinfo  {journal} {Nature Physics}\ }\textbf {\bibinfo {volume}
  {17}},\ \bibinfo {pages} {189} (\bibinfo {year} {2021})}\BibitemShut
  {NoStop}%
\bibitem [{\citenamefont {Cao}\ \emph {et~al.}(2018)\citenamefont {Cao},
  \citenamefont {Fatemi}, \citenamefont {Demir}, \citenamefont {Fang},
  \citenamefont {Tomarken}, \citenamefont {Luo}, \citenamefont
  {Sanchez-Yamagishi}, \citenamefont {Watanabe}, \citenamefont {Taniguchi},
  \citenamefont {Kaxiras}, \citenamefont {Ashoori},\ and\ \citenamefont
  {Jarillo-Herrero}}]{Cao2018}%
  \BibitemOpen
  \bibfield  {author} {\bibinfo {author} {\bibfnamefont {Y.}~\bibnamefont
  {Cao}}, \bibinfo {author} {\bibfnamefont {V.}~\bibnamefont {Fatemi}},
  \bibinfo {author} {\bibfnamefont {A.}~\bibnamefont {Demir}}, \bibinfo
  {author} {\bibfnamefont {S.}~\bibnamefont {Fang}}, \bibinfo {author}
  {\bibfnamefont {S.~L.}\ \bibnamefont {Tomarken}}, \bibinfo {author}
  {\bibfnamefont {J.~Y.}\ \bibnamefont {Luo}}, \bibinfo {author} {\bibfnamefont
  {J.~D.}\ \bibnamefont {Sanchez-Yamagishi}}, \bibinfo {author} {\bibfnamefont
  {K.}~\bibnamefont {Watanabe}}, \bibinfo {author} {\bibfnamefont
  {T.}~\bibnamefont {Taniguchi}}, \bibinfo {author} {\bibfnamefont
  {E.}~\bibnamefont {Kaxiras}}, \bibinfo {author} {\bibfnamefont {R.~C.}\
  \bibnamefont {Ashoori}}, \ and\ \bibinfo {author} {\bibfnamefont
  {P.}~\bibnamefont {Jarillo-Herrero}},\ }\href@noop {} {\bibfield  {journal}
  {\bibinfo  {journal} {Nature}\ }\textbf {\bibinfo {volume} {556}},\ \bibinfo
  {pages} {80} (\bibinfo {year} {2018})}\BibitemShut {NoStop}%
\bibitem [{\citenamefont {Rhim}\ \emph {et~al.}(2020)\citenamefont {Rhim},
  \citenamefont {Kim},\ and\ \citenamefont {Yang}}]{LandauNature2020}%
  \BibitemOpen
  \bibfield  {author} {\bibinfo {author} {\bibfnamefont {J.-W.}\ \bibnamefont
  {Rhim}}, \bibinfo {author} {\bibfnamefont {K.}~\bibnamefont {Kim}}, \ and\
  \bibinfo {author} {\bibfnamefont {B.-J.}\ \bibnamefont {Yang}},\ }\href
  {\doibase 10.1038/s41586-020-2540-1} {\bibfield  {journal} {\bibinfo
  {journal} {Nature}\ }\textbf {\bibinfo {volume} {584}},\ \bibinfo {pages}
  {59} (\bibinfo {year} {2020})}\BibitemShut {NoStop}%
\bibitem [{\citenamefont {Wolf}\ \emph {et~al.}(2019)\citenamefont {Wolf},
  \citenamefont {Lado}, \citenamefont {Blatter},\ and\ \citenamefont
  {Zilberberg}}]{PhysRevLett.123.096802}%
  \BibitemOpen
  \bibfield  {author} {\bibinfo {author} {\bibfnamefont {T.~M.~R.}\
  \bibnamefont {Wolf}}, \bibinfo {author} {\bibfnamefont {J.~L.}\ \bibnamefont
  {Lado}}, \bibinfo {author} {\bibfnamefont {G.}~\bibnamefont {Blatter}}, \
  and\ \bibinfo {author} {\bibfnamefont {O.}~\bibnamefont {Zilberberg}},\
  }\href {\doibase 10.1103/PhysRevLett.123.096802} {\bibfield  {journal}
  {\bibinfo  {journal} {Phys. Rev. Lett.}\ }\textbf {\bibinfo {volume} {123}},\
  \bibinfo {pages} {096802} (\bibinfo {year} {2019})}\BibitemShut {NoStop}%
\bibitem [{\citenamefont {J.}\ \emph {et~al.}(2020)\citenamefont {J.},
  \citenamefont {S.P.},\ and\ \citenamefont {et~al.}}]{Andrei2020}%
  \BibitemOpen
  \bibfield  {author} {\bibinfo {author} {\bibfnamefont {M.}~\bibnamefont
  {J.}}, \bibinfo {author} {\bibfnamefont {M.}~\bibnamefont {S.P.}}, \ and\
  \bibinfo {author} {\bibfnamefont {A.~M.}\ \bibnamefont {et~al.}},\
  }\href@noop {} {\bibfield  {journal} {\bibinfo  {journal} {Nature}\ }\textbf
  {\bibinfo {volume} {584}},\ \bibinfo {pages} {215} (\bibinfo {year}
  {2020})}\BibitemShut {NoStop}%
\bibitem [{\citenamefont {Kempkes}\ \emph {et~al.}(2019)\citenamefont
  {Kempkes}, \citenamefont {Slot}, \citenamefont {Freeney}, \citenamefont
  {Zevenhuizen}, \citenamefont {Vanmaekelbergh}, \citenamefont {Swart},\ and\
  \citenamefont {Smith}}]{nature}%
  \BibitemOpen
  \bibfield  {author} {\bibinfo {author} {\bibfnamefont {S.~N.}\ \bibnamefont
  {Kempkes}}, \bibinfo {author} {\bibfnamefont {M.~R.}\ \bibnamefont {Slot}},
  \bibinfo {author} {\bibfnamefont {S.~E.}\ \bibnamefont {Freeney}}, \bibinfo
  {author} {\bibfnamefont {S.~J.~M.}\ \bibnamefont {Zevenhuizen}}, \bibinfo
  {author} {\bibfnamefont {D.}~\bibnamefont {Vanmaekelbergh}}, \bibinfo
  {author} {\bibfnamefont {I.}~\bibnamefont {Swart}}, \ and\ \bibinfo {author}
  {\bibfnamefont {C.~M.}\ \bibnamefont {Smith}},\ }\href {\doibase
  10.1038/nature26160} {\bibfield  {journal} {\bibinfo  {journal} {Nat. Phys.}\
  }\textbf {\bibinfo {volume} {15}},\ \bibinfo {pages} {127–131} (\bibinfo
  {year} {2019})}\BibitemShut {NoStop}%
\bibitem [{\citenamefont {Manna}\ \emph {et~al.}(2020)\citenamefont {Manna},
  \citenamefont {Pal}, \citenamefont {Wang},\ and\ \citenamefont
  {Nielsen}}]{Biplab2020}%
  \BibitemOpen
  \bibfield  {author} {\bibinfo {author} {\bibfnamefont {S.}~\bibnamefont
  {Manna}}, \bibinfo {author} {\bibfnamefont {B.}~\bibnamefont {Pal}}, \bibinfo
  {author} {\bibfnamefont {W.}~\bibnamefont {Wang}}, \ and\ \bibinfo {author}
  {\bibfnamefont {A.~E.~B.}\ \bibnamefont {Nielsen}},\ }\href {\doibase
  10.1103/PhysRevResearch.2.023401} {\bibfield  {journal} {\bibinfo  {journal}
  {Physical Review Research}\ }\textbf {\bibinfo {volume} {2}},\ \bibinfo
  {pages} {023401} (\bibinfo {year} {2020})}\BibitemShut {NoStop}%
\bibitem [{\citenamefont {Pedersen}(2020)}]{Perdersen2020}%
  \BibitemOpen
  \bibfield  {author} {\bibinfo {author} {\bibfnamefont {T.}~\bibnamefont
  {Pedersen}},\ }\href {\doibase 10.1103/PhysRevB.101.235427} {\bibfield
  {journal} {\bibinfo  {journal} {Phys. Rev. B}\ }\textbf {\bibinfo {volume}
  {101}},\ \bibinfo {pages} {1–7} (\bibinfo {year} {2020})}\BibitemShut
  {NoStop}%
\bibitem [{\citenamefont {Pal}\ and\ \citenamefont {Saha}(2018)}]{Biplab2018}%
  \BibitemOpen
  \bibfield  {author} {\bibinfo {author} {\bibfnamefont {B.}~\bibnamefont
  {Pal}}\ and\ \bibinfo {author} {\bibfnamefont {K.}~\bibnamefont {Saha}},\
  }\href {\doibase 10.1103/PhysRevB.97.195101} {\bibfield  {journal} {\bibinfo
  {journal} {Phys. Rev. B}\ }\textbf {\bibinfo {volume} {97}},\ \bibinfo
  {pages} {195101} (\bibinfo {year} {2018})}\BibitemShut {NoStop}%
\bibitem [{\citenamefont {Leykam}\ \emph {et~al.}(2018)\citenamefont {Leykam},
  \citenamefont {Andreanov},\ and\ \citenamefont {Flach}}]{Reviewflatband2018}%
  \BibitemOpen
  \bibfield  {author} {\bibinfo {author} {\bibfnamefont {D.}~\bibnamefont
  {Leykam}}, \bibinfo {author} {\bibfnamefont {A.}~\bibnamefont {Andreanov}}, \
  and\ \bibinfo {author} {\bibfnamefont {S.}~\bibnamefont {Flach}},\ }\href
  {\doibase 10.1080/23746149.2018.1473052} {\bibfield  {journal} {\bibinfo
  {journal} {Adv. in Phys.}\ }\textbf {\bibinfo {volume} {3}},\ \bibinfo
  {pages} {677} (\bibinfo {year} {2018})}\BibitemShut {NoStop}%
\bibitem [{\citenamefont {Nandy}(2029)}]{Nandy2021}%
  \BibitemOpen
  \bibfield  {author} {\bibinfo {author} {\bibfnamefont {A.}~\bibnamefont
  {Nandy}},\ }\href {\doibase
  https://iopscience.iop.org/article/10.1088/1402-4896/abdcf6} {\bibfield
  {journal} {\bibinfo  {journal} {Physica Scripta}\ }\textbf {\bibinfo {volume}
  {86}},\ \bibinfo {pages} {045802} (\bibinfo {year} {2029})}\BibitemShut
  {NoStop}%
\bibitem [{\citenamefont {Pal}\ and\ \citenamefont
  {Chakrabartia}(2012)}]{Pal2012}%
  \BibitemOpen
  \bibfield  {author} {\bibinfo {author} {\bibfnamefont {B.}~\bibnamefont
  {Pal}}\ and\ \bibinfo {author} {\bibfnamefont {A.}~\bibnamefont
  {Chakrabartia}},\ }\href {\doibase 10.1140/epjb/e2012-30456-8} {\bibfield
  {journal} {\bibinfo  {journal} {Eur. Phys. Journ. B}\ }\textbf {\bibinfo
  {volume} {85}},\ \bibinfo {pages} {307} (\bibinfo {year} {2012})}\BibitemShut
  {NoStop}%
\bibitem [{\citenamefont {Xu}\ \emph {et~al.}(2021)\citenamefont {Xu},
  \citenamefont {Wang}, \citenamefont {Chen}, \citenamefont {Smith},\ and\
  \citenamefont {Jin}}]{XUPHOTONIC}%
  \BibitemOpen
  \bibfield  {author} {\bibinfo {author} {\bibfnamefont {X.-Y.}\ \bibnamefont
  {Xu}}, \bibinfo {author} {\bibfnamefont {X.-W.}\ \bibnamefont {Wang}},
  \bibinfo {author} {\bibfnamefont {D.-Y.}\ \bibnamefont {Chen}}, \bibinfo
  {author} {\bibfnamefont {C.~M.}\ \bibnamefont {Smith}}, \ and\ \bibinfo
  {author} {\bibfnamefont {X.-M.}\ \bibnamefont {Jin}},\ }\href {\doibase
  10.1038/s41566-021-00845-4} {\bibfield  {journal} {\bibinfo  {journal}
  {Nature Photonics}\ }\textbf {\bibinfo {volume} {15}},\ \bibinfo {pages}
  {1749} (\bibinfo {year} {2021})}\BibitemShut {NoStop}%
\bibitem [{\citenamefont {Wang}\ \emph {et~al.}(2018)\citenamefont {Wang},
  \citenamefont {Liu}, \citenamefont {Gu}, \citenamefont {Song}, \citenamefont
  {Wang}, \citenamefont {Jiang}, \citenamefont {Zhang}, \citenamefont {Han},
  \citenamefont {Hao}, \citenamefont {Bai}, \citenamefont {Wang}, \citenamefont
  {Li}, \citenamefont {Xu},\ and\ \citenamefont {Li}}]{Wang2018}%
  \BibitemOpen
  \bibfield  {author} {\bibinfo {author} {\bibfnamefont {L.}~\bibnamefont
  {Wang}}, \bibinfo {author} {\bibfnamefont {R.}~\bibnamefont {Liu}}, \bibinfo
  {author} {\bibfnamefont {J.}~\bibnamefont {Gu}}, \bibinfo {author}
  {\bibfnamefont {B.}~\bibnamefont {Song}}, \bibinfo {author} {\bibfnamefont
  {H.}~\bibnamefont {Wang}}, \bibinfo {author} {\bibfnamefont {X.}~\bibnamefont
  {Jiang}}, \bibinfo {author} {\bibfnamefont {K.}~\bibnamefont {Zhang}},
  \bibinfo {author} {\bibfnamefont {X.}~\bibnamefont {Han}}, \bibinfo {author}
  {\bibfnamefont {X.-Q.}\ \bibnamefont {Hao}}, \bibinfo {author} {\bibfnamefont
  {S.}~\bibnamefont {Bai}}, \bibinfo {author} {\bibfnamefont {M.}~\bibnamefont
  {Wang}}, \bibinfo {author} {\bibfnamefont {X.}~\bibnamefont {Li}}, \bibinfo
  {author} {\bibfnamefont {B.}~\bibnamefont {Xu}}, \ and\ \bibinfo {author}
  {\bibfnamefont {X.}~\bibnamefont {Li}},\ }\href {\doibase
  10.1021/jacs.8b05530} {\bibfield  {journal} {\bibinfo  {journal} {Journal of
  the American Chemical Society}\ }\textbf {\bibinfo {volume} {140}},\ \bibinfo
  {pages} {14087} (\bibinfo {year} {2018})}\BibitemShut {NoStop}%
\bibitem [{\citenamefont {Zhang}\ \emph {et~al.}(2018)\citenamefont {Zhang},
  \citenamefont {Gu}, \citenamefont {Li}, \citenamefont {Wang}, \citenamefont
  {Tang}, \citenamefont {Zhang}, \citenamefont {Hou},\ and\ \citenamefont
  {Wang}}]{Chain2018}%
  \BibitemOpen
  \bibfield  {author} {\bibinfo {author} {\bibfnamefont {X.}~\bibnamefont
  {Zhang}}, \bibinfo {author} {\bibfnamefont {G.}~\bibnamefont {Gu}}, \bibinfo
  {author} {\bibfnamefont {N.}~\bibnamefont {Li}}, \bibinfo {author}
  {\bibfnamefont {H.}~\bibnamefont {Wang}}, \bibinfo {author} {\bibfnamefont
  {H.}~\bibnamefont {Tang}}, \bibinfo {author} {\bibfnamefont {Y.}~\bibnamefont
  {Zhang}}, \bibinfo {author} {\bibfnamefont {S.}~\bibnamefont {Hou}}, \ and\
  \bibinfo {author} {\bibfnamefont {Y.}~\bibnamefont {Wang}},\ }\href {\doibase
  10.1039/c7ra11825b} {\bibfield  {journal} {\bibinfo  {journal} {RSC Adv.}\
  }\textbf {\bibinfo {volume} {8}},\ \bibinfo {pages} {1852} (\bibinfo {year}
  {2018})}\BibitemShut {NoStop}%
\bibitem [{\citenamefont {Jiang}\ \emph {et~al.}(2020)\citenamefont {Jiang},
  \citenamefont {Liu}, \citenamefont {Chen}, \citenamefont {Wang},
  \citenamefont {Zhao}, \citenamefont {Li}, \citenamefont {Zhang},
  \citenamefont {Xie}, \citenamefont {Wang}, \citenamefont {Li}, \citenamefont
  {Newkome},\ and\ \citenamefont {Wang}}]{JIANG2020101064}%
  \BibitemOpen
  \bibfield  {author} {\bibinfo {author} {\bibfnamefont {Z.}~\bibnamefont
  {Jiang}}, \bibinfo {author} {\bibfnamefont {D.}~\bibnamefont {Liu}}, \bibinfo
  {author} {\bibfnamefont {M.}~\bibnamefont {Chen}}, \bibinfo {author}
  {\bibfnamefont {J.}~\bibnamefont {Wang}}, \bibinfo {author} {\bibfnamefont
  {H.}~\bibnamefont {Zhao}}, \bibinfo {author} {\bibfnamefont {Y.}~\bibnamefont
  {Li}}, \bibinfo {author} {\bibfnamefont {Z.}~\bibnamefont {Zhang}}, \bibinfo
  {author} {\bibfnamefont {T.}~\bibnamefont {Xie}}, \bibinfo {author}
  {\bibfnamefont {F.}~\bibnamefont {Wang}}, \bibinfo {author} {\bibfnamefont
  {X.}~\bibnamefont {Li}}, \bibinfo {author} {\bibfnamefont {G.~R.}\
  \bibnamefont {Newkome}}, \ and\ \bibinfo {author} {\bibfnamefont
  {P.}~\bibnamefont {Wang}},\ }\href {\doibase
  https://doi.org/10.1016/j.isci.2020.101064} {\bibfield  {journal} {\bibinfo
  {journal} {Science}\ }\textbf {\bibinfo {volume} {23}},\ \bibinfo {pages}
  {101064} (\bibinfo {year} {2020})}\BibitemShut {NoStop}%
\bibitem [{\citenamefont {Rastgoo-Lahrood}\ \emph {et~al.}(2016)\citenamefont
  {Rastgoo-Lahrood}, \citenamefont {Martsinovich}, \citenamefont {Lischka},
  \citenamefont {Eichhorn}, \citenamefont {Szabelski}, \citenamefont
  {Nieckarz}, \citenamefont {Strunskus}, \citenamefont {Das}, \citenamefont
  {Schmittel}, \citenamefont {Heckl},\ and\ \citenamefont
  {Lackinger}}]{Rastgoo}%
  \BibitemOpen
  \bibfield  {author} {\bibinfo {author} {\bibfnamefont {A.}~\bibnamefont
  {Rastgoo-Lahrood}}, \bibinfo {author} {\bibfnamefont {N.}~\bibnamefont
  {Martsinovich}}, \bibinfo {author} {\bibfnamefont {M.}~\bibnamefont
  {Lischka}}, \bibinfo {author} {\bibfnamefont {J.}~\bibnamefont {Eichhorn}},
  \bibinfo {author} {\bibfnamefont {P.}~\bibnamefont {Szabelski}}, \bibinfo
  {author} {\bibfnamefont {D.}~\bibnamefont {Nieckarz}}, \bibinfo {author}
  {\bibfnamefont {T.}~\bibnamefont {Strunskus}}, \bibinfo {author}
  {\bibfnamefont {K.}~\bibnamefont {Das}}, \bibinfo {author} {\bibfnamefont
  {M.}~\bibnamefont {Schmittel}}, \bibinfo {author} {\bibfnamefont {W.~M.}\
  \bibnamefont {Heckl}}, \ and\ \bibinfo {author} {\bibfnamefont
  {M.}~\bibnamefont {Lackinger}},\ }\href {\doibase 10.1021/acsnano.6b05470}
  {\bibfield  {journal} {\bibinfo  {journal} {ACS Nano}\ }\textbf {\bibinfo
  {volume} {10}},\ \bibinfo {pages} {12} (\bibinfo {year} {2016})}\BibitemShut
  {NoStop}%
\bibitem [{\citenamefont {Jian}\ \emph {et~al.}(2015)\citenamefont {Jian},
  \citenamefont {Yongfeng}, \citenamefont {Min}, \citenamefont {Jingxin},
  \citenamefont {Xiong}, \citenamefont {Julian}, \citenamefont {Gerhard},
  \citenamefont {Xiang}, \citenamefont {Michael},\ and\ \citenamefont
  {Kai}}]{JIANNATURE}%
  \BibitemOpen
  \bibfield  {author} {\bibinfo {author} {\bibfnamefont {S.}~\bibnamefont
  {Jian}}, \bibinfo {author} {\bibfnamefont {W.}~\bibnamefont {Yongfeng}},
  \bibinfo {author} {\bibfnamefont {C.}~\bibnamefont {Min}}, \bibinfo {author}
  {\bibfnamefont {D.}~\bibnamefont {Jingxin}}, \bibinfo {author} {\bibfnamefont
  {Z.}~\bibnamefont {Xiong}}, \bibinfo {author} {\bibfnamefont
  {K.}~\bibnamefont {Julian}}, \bibinfo {author} {\bibfnamefont
  {H.}~\bibnamefont {Gerhard}}, \bibinfo {author} {\bibfnamefont
  {S.}~\bibnamefont {Xiang}}, \bibinfo {author} {\bibfnamefont {G.~J.}\
  \bibnamefont {Michael}}, \ and\ \bibinfo {author} {\bibfnamefont
  {W.}~\bibnamefont {Kai}},\ }\href {\doibase 10.1038/nchem.2211} {\bibfield
  {journal} {\bibinfo  {journal} {Nature Chemistry}\ }\textbf {\bibinfo
  {volume} {7}},\ \bibinfo {pages} {389} (\bibinfo {year} {2015})}\BibitemShut
  {NoStop}%
\bibitem [{\citenamefont {Zhang}\ \emph {et~al.}(2015)\citenamefont {Zhang},
  \citenamefont {Li}, \citenamefont {Gu}, \citenamefont {Wang}, \citenamefont
  {Nieckarz}, \citenamefont {Szabelski}, \citenamefont {He}, \citenamefont
  {Wang}, \citenamefont {Xie}, \citenamefont {Shen}, \citenamefont {Lü},
  \citenamefont {Tang}, \citenamefont {Peng}, \citenamefont {Hou},
  \citenamefont {Wu},\ and\ \citenamefont {Wang}}]{Zhang2}%
  \BibitemOpen
  \bibfield  {author} {\bibinfo {author} {\bibfnamefont {X.}~\bibnamefont
  {Zhang}}, \bibinfo {author} {\bibfnamefont {N.}~\bibnamefont {Li}}, \bibinfo
  {author} {\bibfnamefont {G.-C.}\ \bibnamefont {Gu}}, \bibinfo {author}
  {\bibfnamefont {H.}~\bibnamefont {Wang}}, \bibinfo {author} {\bibfnamefont
  {D.}~\bibnamefont {Nieckarz}}, \bibinfo {author} {\bibfnamefont
  {P.}~\bibnamefont {Szabelski}}, \bibinfo {author} {\bibfnamefont
  {Y.}~\bibnamefont {He}}, \bibinfo {author} {\bibfnamefont {Y.}~\bibnamefont
  {Wang}}, \bibinfo {author} {\bibfnamefont {C.}~\bibnamefont {Xie}}, \bibinfo
  {author} {\bibfnamefont {Z.-Y.}\ \bibnamefont {Shen}}, \bibinfo {author}
  {\bibfnamefont {J.-T.}\ \bibnamefont {Lü}}, \bibinfo {author} {\bibfnamefont
  {H.}~\bibnamefont {Tang}}, \bibinfo {author} {\bibfnamefont {L.-M.}\
  \bibnamefont {Peng}}, \bibinfo {author} {\bibfnamefont {S.-M.}\ \bibnamefont
  {Hou}}, \bibinfo {author} {\bibfnamefont {K.}~\bibnamefont {Wu}}, \ and\
  \bibinfo {author} {\bibfnamefont {Y.-F.}\ \bibnamefont {Wang}},\ }\href@noop
  {} {\bibfield  {journal} {\bibinfo  {journal} {ACS Nano}\ }\textbf {\bibinfo
  {volume} {9}},\ \bibinfo {pages} {11909} (\bibinfo {year}
  {2015})}\BibitemShut {NoStop}%
\bibitem [{\citenamefont {Li}\ \emph {et~al.}(2017{\natexlab{a}})\citenamefont
  {Li}, \citenamefont {Gu}, \citenamefont {Zhang}, \citenamefont {Song},
  \citenamefont {Zhang}, \citenamefont {Teo}, \citenamefont {Peng},
  \citenamefont {Hou},\ and\ \citenamefont {Wang}}]{templating1}%
  \BibitemOpen
  \bibfield  {author} {\bibinfo {author} {\bibfnamefont {N.}~\bibnamefont
  {Li}}, \bibinfo {author} {\bibfnamefont {G.}~\bibnamefont {Gu}}, \bibinfo
  {author} {\bibfnamefont {X.}~\bibnamefont {Zhang}}, \bibinfo {author}
  {\bibfnamefont {D.}~\bibnamefont {Song}}, \bibinfo {author} {\bibfnamefont
  {Y.}~\bibnamefont {Zhang}}, \bibinfo {author} {\bibfnamefont {B.~K.}\
  \bibnamefont {Teo}}, \bibinfo {author} {\bibfnamefont {L.-M.}\ \bibnamefont
  {Peng}}, \bibinfo {author} {\bibfnamefont {S.}~\bibnamefont {Hou}}, \ and\
  \bibinfo {author} {\bibfnamefont {Y.}~\bibnamefont {Wang}},\ }\href {\doibase
  10.1039/C7CC00566K} {\bibfield  {journal} {\bibinfo  {journal} {Chem.
  Commun.}\ }\textbf {\bibinfo {volume} {53}},\ \bibinfo {pages} {3469}
  (\bibinfo {year} {2017}{\natexlab{a}})}\BibitemShut {NoStop}%
\bibitem [{\citenamefont {Li}\ \emph {et~al.}(2015)\citenamefont {Li},
  \citenamefont {Zhang}, \citenamefont {Gu}, \citenamefont {Wang},
  \citenamefont {Nieckarz}, \citenamefont {Szabelski}, \citenamefont {He},
  \citenamefont {Wang}, \citenamefont {Lü}, \citenamefont {Tang},
  \citenamefont {Peng}, \citenamefont {Hou}, \citenamefont {Wu},\ and\
  \citenamefont {Wang}}]{LI20151198}%
  \BibitemOpen
  \bibfield  {author} {\bibinfo {author} {\bibfnamefont {N.}~\bibnamefont
  {Li}}, \bibinfo {author} {\bibfnamefont {X.}~\bibnamefont {Zhang}}, \bibinfo
  {author} {\bibfnamefont {G.-C.}\ \bibnamefont {Gu}}, \bibinfo {author}
  {\bibfnamefont {H.}~\bibnamefont {Wang}}, \bibinfo {author} {\bibfnamefont
  {D.}~\bibnamefont {Nieckarz}}, \bibinfo {author} {\bibfnamefont
  {P.}~\bibnamefont {Szabelski}}, \bibinfo {author} {\bibfnamefont
  {Y.}~\bibnamefont {He}}, \bibinfo {author} {\bibfnamefont {Y.}~\bibnamefont
  {Wang}}, \bibinfo {author} {\bibfnamefont {J.-T.}\ \bibnamefont {Lü}},
  \bibinfo {author} {\bibfnamefont {H.}~\bibnamefont {Tang}}, \bibinfo {author}
  {\bibfnamefont {L.-M.}\ \bibnamefont {Peng}}, \bibinfo {author}
  {\bibfnamefont {S.-M.}\ \bibnamefont {Hou}}, \bibinfo {author} {\bibfnamefont
  {K.}~\bibnamefont {Wu}}, \ and\ \bibinfo {author} {\bibfnamefont {Y.-F.}\
  \bibnamefont {Wang}},\ }\href {\doibase
  https://doi.org/10.1016/j.cclet.2015.08.006} {\bibfield  {journal} {\bibinfo
  {journal} {Chinese Chemical Letters}\ }\textbf {\bibinfo {volume} {26}},\
  \bibinfo {pages} {1198} (\bibinfo {year} {2015})}\BibitemShut {NoStop}%
\bibitem [{\citenamefont {Li}\ \emph {et~al.}(2017{\natexlab{b}})\citenamefont
  {Li}, \citenamefont {Gu}, \citenamefont {Zhang}, \citenamefont {Song},
  \citenamefont {Zhang}, \citenamefont {Teo}, \citenamefont {Peng},
  \citenamefont {Hou},\ and\ \citenamefont {Wang}}]{C7CC00566K}%
  \BibitemOpen
  \bibfield  {author} {\bibinfo {author} {\bibfnamefont {N.}~\bibnamefont
  {Li}}, \bibinfo {author} {\bibfnamefont {G.}~\bibnamefont {Gu}}, \bibinfo
  {author} {\bibfnamefont {X.}~\bibnamefont {Zhang}}, \bibinfo {author}
  {\bibfnamefont {D.}~\bibnamefont {Song}}, \bibinfo {author} {\bibfnamefont
  {Y.}~\bibnamefont {Zhang}}, \bibinfo {author} {\bibfnamefont {B.~K.}\
  \bibnamefont {Teo}}, \bibinfo {author} {\bibfnamefont {L.-M.}\ \bibnamefont
  {Peng}}, \bibinfo {author} {\bibfnamefont {S.}~\bibnamefont {Hou}}, \ and\
  \bibinfo {author} {\bibfnamefont {Y.}~\bibnamefont {Wang}},\ }\href {\doibase
  10.1039/C7CC00566K} {\bibfield  {journal} {\bibinfo  {journal} {Chem.
  Commun.}\ }\textbf {\bibinfo {volume} {53}},\ \bibinfo {pages} {3469}
  (\bibinfo {year} {2017}{\natexlab{b}})}\BibitemShut {NoStop}%
\bibitem [{\citenamefont {Zhang}\ \emph {et~al.}(2020)\citenamefont {Zhang},
  \citenamefont {Zhang}, \citenamefont {Li}, \citenamefont {Zhao},
  \citenamefont {Hou}, \citenamefont {Wu},\ and\ \citenamefont
  {Wang}}]{2dcrystal2020}%
  \BibitemOpen
  \bibfield  {author} {\bibinfo {author} {\bibfnamefont {Y.}~\bibnamefont
  {Zhang}}, \bibinfo {author} {\bibfnamefont {X.}~\bibnamefont {Zhang}},
  \bibinfo {author} {\bibfnamefont {Y.}~\bibnamefont {Li}}, \bibinfo {author}
  {\bibfnamefont {S.}~\bibnamefont {Zhao}}, \bibinfo {author} {\bibfnamefont
  {S.}~\bibnamefont {Hou}}, \bibinfo {author} {\bibfnamefont {K.}~\bibnamefont
  {Wu}}, \ and\ \bibinfo {author} {\bibfnamefont {Y.}~\bibnamefont {Wang}},\
  }\href {\doibase 10.1021/jacs.0c08979} {\bibfield  {journal} {\bibinfo
  {journal} {J. Am. Chem. Soc.}\ }\textbf {\bibinfo {volume} {142}},\ \bibinfo
  {pages} {17928} (\bibinfo {year} {2020})}\BibitemShut {NoStop}%
\bibitem [{\citenamefont {Datta}(1995)}]{Data}%
  \BibitemOpen
  \bibfield  {author} {\bibinfo {author} {\bibfnamefont {S.}~\bibnamefont
  {Datta}},\ }\href@noop {} {\emph {\bibinfo {title} {In Electronic Transport
  in Mesoscopic Systems}}}\ (\bibinfo  {publisher} {Cambridge},\ \bibinfo
  {year} {1995})\BibitemShut {NoStop}%
\bibitem [{\citenamefont {Santos}\ \emph {et~al.}(2020)\citenamefont {Santos},
  \citenamefont {Latgé}, \citenamefont {Brey},\ and\ \citenamefont
  {Chico}}]{Carbon2020}%
  \BibitemOpen
  \bibfield  {author} {\bibinfo {author} {\bibfnamefont {H.}~\bibnamefont
  {Santos}}, \bibinfo {author} {\bibfnamefont {A.}~\bibnamefont {Latgé}},
  \bibinfo {author} {\bibfnamefont {L.}~\bibnamefont {Brey}}, \ and\ \bibinfo
  {author} {\bibfnamefont {L.}~\bibnamefont {Chico}},\ }\href {\doibase
  10.1016/j.carbon.2020.05.054} {\bibfield  {journal} {\bibinfo  {journal}
  {Carbon}\ }\textbf {\bibinfo {volume} {168}},\ \bibinfo {pages} {1} (\bibinfo
  {year} {2020})}\BibitemShut {NoStop}%
\bibitem [{\citenamefont {Chico}\ \emph
  {et~al.}(2015{\natexlab{a}})\citenamefont {Chico}, \citenamefont {Latgé},\
  and\ \citenamefont {Brey}}]{Leonor2015}%
  \BibitemOpen
  \bibfield  {author} {\bibinfo {author} {\bibfnamefont {L.}~\bibnamefont
  {Chico}}, \bibinfo {author} {\bibfnamefont {A.}~\bibnamefont {Latgé}}, \
  and\ \bibinfo {author} {\bibfnamefont {L.}~\bibnamefont {Brey}},\ }\href
  {\doibase 10.1039/C5CP01637A} {\bibfield  {journal} {\bibinfo  {journal}
  {Phys. Chem. Chem. Phys.}\ }\textbf {\bibinfo {volume} {17}},\ \bibinfo
  {pages} {16469} (\bibinfo {year} {2015}{\natexlab{a}})}\BibitemShut {NoStop}%
\bibitem [{\citenamefont {Wang}(1995)}]{Wang1995}%
  \BibitemOpen
  \bibfield  {author} {\bibinfo {author} {\bibfnamefont {X.~R.}\ \bibnamefont
  {Wang}},\ }\href {\doibase 10.1103/PhysRevB.51.9310} {\bibfield  {journal}
  {\bibinfo  {journal} {Phys. Rev. B}\ }\textbf {\bibinfo {volume} {51}},\
  \bibinfo {pages} {9310} (\bibinfo {year} {1995})}\BibitemShut {NoStop}%
\bibitem [{\citenamefont {Silveira}\ \emph {et~al.}(2016)\citenamefont
  {Silveira}, \citenamefont {Alexandre},\ and\ \citenamefont
  {Chacham}}]{Chachan2016}%
  \BibitemOpen
  \bibfield  {author} {\bibinfo {author} {\bibfnamefont {O.~J.}\ \bibnamefont
  {Silveira}}, \bibinfo {author} {\bibfnamefont {S.~S.}\ \bibnamefont
  {Alexandre}}, \ and\ \bibinfo {author} {\bibfnamefont {H.}~\bibnamefont
  {Chacham}},\ }\href {\doibase 10.1021/acs.jpcc.6b05081} {\bibfield  {journal}
  {\bibinfo  {journal} {J. Phys. Chem. C}\ }\textbf {\bibinfo {volume} {120}},\
  \bibinfo {pages} {19796} (\bibinfo {year} {2016})}\BibitemShut {NoStop}%
\bibitem [{\citenamefont {Foroutan-pour}\ \emph {et~al.}(1999)\citenamefont
  {Foroutan-pour}, \citenamefont {Dutilleul},\ and\ \citenamefont
  {Smith}}]{FOROUTANPOUR1999195}%
  \BibitemOpen
  \bibfield  {author} {\bibinfo {author} {\bibfnamefont {K.}~\bibnamefont
  {Foroutan-pour}}, \bibinfo {author} {\bibfnamefont {P.}~\bibnamefont
  {Dutilleul}}, \ and\ \bibinfo {author} {\bibfnamefont {D.}~\bibnamefont
  {Smith}},\ }\href {\doibase https://doi.org/10.1016/S0096-3003(98)10096-6}
  {\bibfield  {journal} {\bibinfo  {journal} {Applied Mathematics and
  Computation}\ }\textbf {\bibinfo {volume} {105}},\ \bibinfo {pages} {195}
  (\bibinfo {year} {1999})}\BibitemShut {NoStop}%
\bibitem [{\citenamefont {Kaurov}(2012)}]{forum}%
  \BibitemOpen
  \bibfield  {author} {\bibinfo {author} {\bibfnamefont {V.}~\bibnamefont
  {Kaurov}},\ }\href
  {https://www.mathematica.stackexchange.com/questions/13125/measuring-fractal-dimension-of-natural-objects-from-digital-images}
  {\bibfield  {journal} {\bibinfo  {journal} {Measuring Fractal Dimension of
  Natural Object from Digital Images, Mathematica Stack Exchange}\ } (\bibinfo
  {year} {2012})}\BibitemShut {NoStop}%
\bibitem [{\citenamefont {Rosales}\ \emph {et~al.}(2008)\citenamefont
  {Rosales}, \citenamefont {Pacheco}, \citenamefont {Barticevic}, \citenamefont
  {Latg{\'{e}}},\ and\ \citenamefont {Orellana}}]{Latge2008}%
  \BibitemOpen
  \bibfield  {author} {\bibinfo {author} {\bibfnamefont {L.}~\bibnamefont
  {Rosales}}, \bibinfo {author} {\bibfnamefont {M.}~\bibnamefont {Pacheco}},
  \bibinfo {author} {\bibfnamefont {Z.}~\bibnamefont {Barticevic}}, \bibinfo
  {author} {\bibfnamefont {A.}~\bibnamefont {Latg{\'{e}}}}, \ and\ \bibinfo
  {author} {\bibfnamefont {P.~A.}\ \bibnamefont {Orellana}},\ }\href {\doibase
  10.1088/0957-4484/19/6/065402} {\bibfield  {journal} {\bibinfo  {journal}
  {Nanotechnology}\ }\textbf {\bibinfo {volume} {19}},\ \bibinfo {pages}
  {065402} (\bibinfo {year} {2008})}\BibitemShut {NoStop}%
\bibitem [{\citenamefont {Chico}\ \emph
  {et~al.}(2015{\natexlab{b}})\citenamefont {Chico}, \citenamefont {Latgé},\
  and\ \citenamefont {Brey}}]{Latge2015}%
  \BibitemOpen
  \bibfield  {author} {\bibinfo {author} {\bibfnamefont {L.}~\bibnamefont
  {Chico}}, \bibinfo {author} {\bibfnamefont {A.}~\bibnamefont {Latgé}}, \
  and\ \bibinfo {author} {\bibfnamefont {L.}~\bibnamefont {Brey}},\ }\href
  {\doibase https://doi.org/10.1039/C5CP01637A} {\bibfield  {journal} {\bibinfo
   {journal} {Phys. Chem. Chem. Phys.}\ }\textbf {\bibinfo {volume} {17}},\
  \bibinfo {pages} {16469} (\bibinfo {year} {2015}{\natexlab{b}})}\BibitemShut
  {NoStop}%
\bibitem [{\citenamefont {Torres}\ \emph {et~al.}(2021)\citenamefont {Torres},
  \citenamefont {Faria},\ and\ \citenamefont {Latg\'e}}]{Latge2021}%
  \BibitemOpen
  \bibfield  {author} {\bibinfo {author} {\bibfnamefont {V.}~\bibnamefont
  {Torres}}, \bibinfo {author} {\bibfnamefont {D.}~\bibnamefont {Faria}}, \
  and\ \bibinfo {author} {\bibfnamefont {A.}~\bibnamefont {Latg\'e}},\ }\href
  {\doibase 10.1103/PhysRevB.103.115437} {\bibfield  {journal} {\bibinfo
  {journal} {Phys. Rev. B}\ }\textbf {\bibinfo {volume} {103}},\ \bibinfo
  {pages} {115437} (\bibinfo {year} {2021})}\BibitemShut {NoStop}%
\bibitem [{\citenamefont {Li}\ \emph {et~al.}(2017{\natexlab{c}})\citenamefont
  {Li}, \citenamefont {Gu}, \citenamefont {Zhang}, \citenamefont {Song},
  \citenamefont {Zhang}, \citenamefont {Teo}, \citenamefont {Peng},
  \citenamefont {Hou},\ and\ \citenamefont {Wang}}]{Liu2017}%
  \BibitemOpen
  \bibfield  {author} {\bibinfo {author} {\bibfnamefont {N.}~\bibnamefont
  {Li}}, \bibinfo {author} {\bibfnamefont {G.}~\bibnamefont {Gu}}, \bibinfo
  {author} {\bibfnamefont {X.}~\bibnamefont {Zhang}}, \bibinfo {author}
  {\bibfnamefont {D.}~\bibnamefont {Song}}, \bibinfo {author} {\bibfnamefont
  {Y.}~\bibnamefont {Zhang}}, \bibinfo {author} {\bibfnamefont {B.~K.}\
  \bibnamefont {Teo}}, \bibinfo {author} {\bibfnamefont {L.-m.}\ \bibnamefont
  {Peng}}, \bibinfo {author} {\bibfnamefont {S.}~\bibnamefont {Hou}}, \ and\
  \bibinfo {author} {\bibfnamefont {Y.}~\bibnamefont {Wang}},\ }\href {\doibase
  10.1039/C7CC00566K} {\bibfield  {journal} {\bibinfo  {journal} {Chem.
  Commun.}\ }\textbf {\bibinfo {volume} {53}},\ \bibinfo {pages} {3469}
  (\bibinfo {year} {2017}{\natexlab{c}})}\BibitemShut {NoStop}%
\bibitem [{\citenamefont {Maciá}\ and\ \citenamefont {Adame}(1996)}]{macia}%
  \BibitemOpen
  \bibfield  {author} {\bibinfo {author} {\bibfnamefont {E.}~\bibnamefont
  {Maciá}}\ and\ \bibinfo {author} {\bibfnamefont {F.~D.}\ \bibnamefont
  {Adame}},\ }\href@noop {} {\bibfield  {journal} {\bibinfo  {journal}
  {Semicond. Sci. Technol.}\ }\textbf {\bibinfo {volume} {11}},\ \bibinfo
  {pages} {1041} (\bibinfo {year} {1996})}\BibitemShut {NoStop}%
\end{thebibliography}%

\end{document}